\documentclass[aps,prb,twocolumn,superscriptaddress]{revtex4-1}

\usepackage{natbib}

\usepackage{amsmath}

\usepackage{amsfonts}

\usepackage[makeroom]{cancel}

\usepackage{graphicx}

\usepackage{mathrsfs}

\usepackage{euscript}

\usepackage{color}

\begin{document}


\title{Exact solutions of Kondo problems in higher-order fermions}


\author{Peng Song}
\affiliation{Department of Physics, Nanjing University, Nanjing 210093, China}
\affiliation{National Laboratory of Solid State Microstructures and Collaborative Innovation Center of Advanced Microstructures, Nanjing University, Nanjing 210093, China}
\author{Haodong Ma}
\affiliation{Department of Physics, Nanjing University, Nanjing 210093, China}
\affiliation{National Laboratory of Solid State Microstructures and Collaborative Innovation Center of Advanced Microstructures, Nanjing University, Nanjing 210093, China}
\author{Rui Wang}
\email{rwang89@nju.edu.cn}
\affiliation{Department of Physics, Nanjing University, Nanjing 210093, China}
\affiliation{National Laboratory of Solid State Microstructures and Collaborative Innovation Center of Advanced Microstructures, Nanjing University, Nanjing 210093, China}
\author{Baigeng Wang}
\email{bgwang@nju.edu.cn}
\affiliation{Department of Physics, Nanjing University, Nanjing 210093, China}
\affiliation{National Laboratory of Solid State Microstructures and Collaborative Innovation Center of Advanced Microstructures, Nanjing University, Nanjing 210093, China}



\begin{abstract}
The conformal field theory (CFT) approach to Kondo problems, originally developed by Affleck and Ludwig (AL),  has greatly advanced the fundamental knowledge of Kondo physics.  The CFT approach to Kondo impurities is based on a necessary approximation, i.e.,  the linearization of the low-lying excitations in a narrow energy window about the Fermi surface. This treatment works well in normal metal baths, but encounters fundamental difficulties in systems with Fermi points and high-order dispersion relations. Prominent examples of such systems are the recently-proposed topological semimetals with emergent higher-order fermions. Here, we develop a new CFT technique that yields exact solutions to the Kondo problems in higher-order fermion systems. Our approach does not require any linearization of the low-lying excitations, and more importantly, it rigorously bosonizes the entire energy spectrum of the higher-order fermions. Therefore, it provides a more solid theoretical base for evaluating the thermodynamic quantities  at finite temperatures.  Our work significantly broadens the  scope of CFT techniques and brings about unprecedented applications beyond the reach of conventional methods.
\end{abstract}


\maketitle

\section{Introduction}
\begin{figure}
\includegraphics[width=\linewidth]{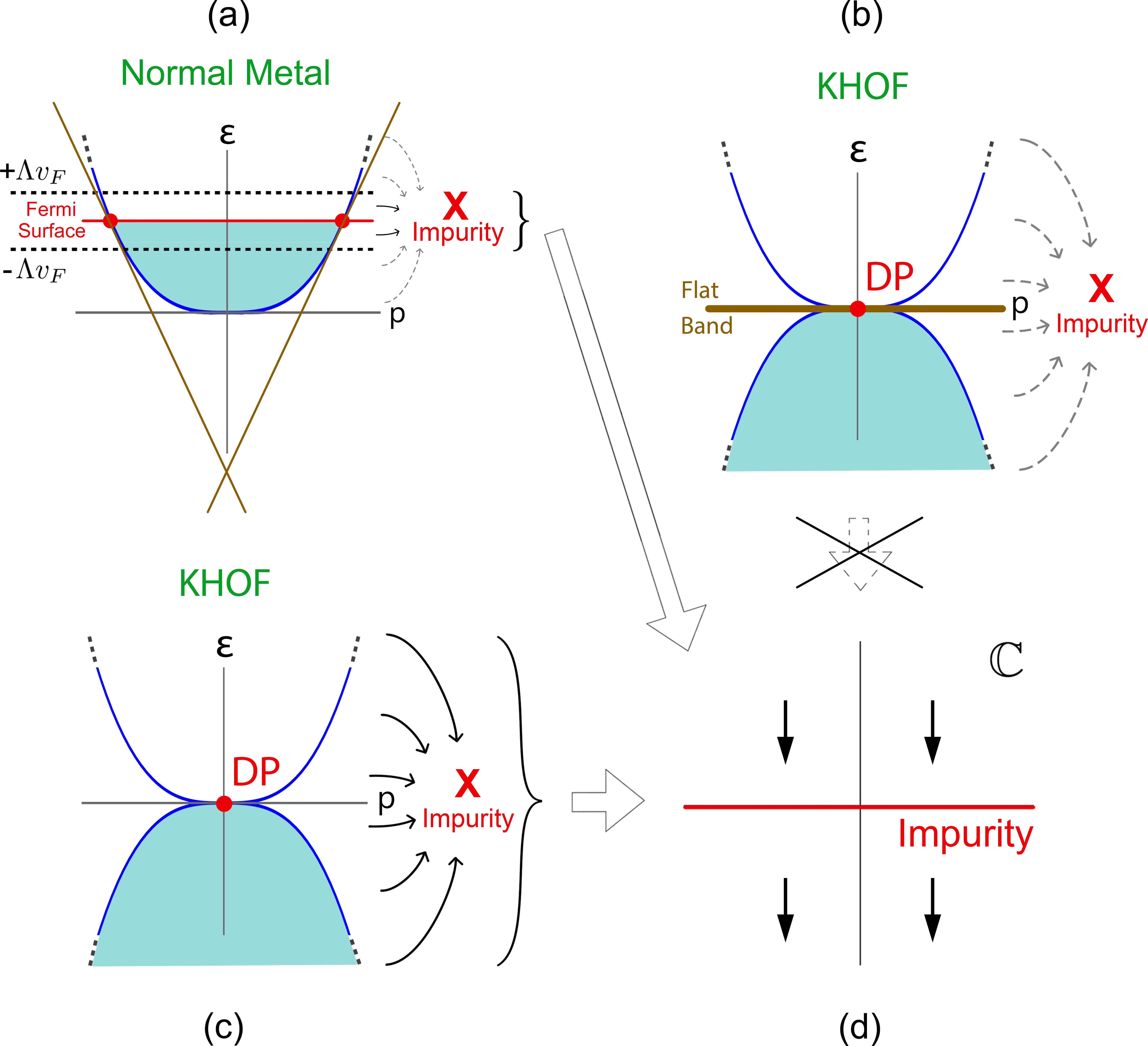}
\caption{\label{fig:MappingToComplexPlane} Figure (a) denotes AL's original approach applied to normal metals, in which an artificial cut-off $\pm \Lambda v_F$ about the Fermi surface has to be introduced. Only low-lying excitations satisfying $-\Lambda v_F < \epsilon < +\Lambda v_F$, which are approximated to be linear, are mapped into the complex plane denoted by Figure (d), although the impurity also couples to the higher-energy regimes of the bath. Figure (b) shows how AL's original approach becomes inapplicable to KHOF systems, which have their Fermi points located exactly at their Dirac points. AL's cut-off becomes infinitesimally narrow and their linearization approximation produces a flat band; no degrees of freedom in the KHOF model can be mapped into the complex plane. Figure (c) denotes our FEMCFT approach applied to KHOF systems. No energy cut-offs and linearization approximations are required, and the full energy spectrum is mapped into the complex plane, establishing an exact CFT solution to KHOF models.}
\end{figure}

The Kondo effect \cite{KondoOriginal}, which treats a quantum magnetic impurity in normal metals, invoked fundamental developments in condensed matter physics. Among these progresses, the conformal field theory (CFT) approach developed by Affleck and Ludwig (AL) unveiled the physical nature of the Kondo problem, and demonstrated the elegant connections between conformal symmetry and the Kondo fixed point \cite{AL1,AL2,AL3,AL4,AL5,AL6,AL7,VonDelft}. In their original approach, AL mapped the multi-channel Kondo problem to an exactly-solvable CFT in the complex plane. In order to establish the mapping, they linearized the low-lying excitations in a narrow energy window $-\Lambda v_F < \epsilon < +\Lambda v_F$ about the Fermi surface, with $v_F$ being the Fermi velocity and $\Lambda$ being an artificially introduced cut-off, and ignored the higher-energy excitations, as a necessary approximation (Fig.~\ref{fig:MappingToComplexPlane}(a)). Although this treatment works well for normal metals with well-defined Fermi surfaces, it will encounter difficulties in more exotic thermals baths, including the pseudogapped systems \cite{BFKM} and those with Fermi points, such as topological semimetals.

On the other hand, past years have witnessed increasing interests in the Kondo problem in topological materials with emergent particles, including graphene, Weyl and Dirac semimetals. In particular, most recently, even more exotic Weyl and Dirac semimetals which exhibit effective fermions with higher-order dispersion relations have been proposed. These higher-order dispersion relations take a form that is linear in one direction, but quadratic or cubic in the orthogonal plane \cite{HigherOrder_1, HigherOrder_2, HigherOrder_3, HigherOrder_4, HigherOrder_5, HigherOrder_6, HigherOrder_7, HigherOrder_8, HigherOrder_9, HigherOrder_10, HigherOrder_11, HigherOrder_12, HigherOrder_13, HigherOrder_14, HigherOrder_15}. Such materials are named quadratic and cubic Weyl/Dirac semimetals respectively \cite{HigherOrder_2, HigherOrder_7, HigherOrder_12}, with their corresponding emergent fermions referred to as quadratic and cubic Weyl/Dirac fermions respectively \cite{HigherOrder_1, HigherOrder_2, HigherOrder_5}, and they are classified as Weyl/Dirac according to their 2-fold/4-fold band degeneracies at their band crossings \cite{HigherOrder_2, HigherOrder_3, HigherOrder_5}. For example, in three dimensions, the cubic Weyl/Dirac semimetals display the dispersion relation, in its simplest form,
\begin{equation}\label{AnisotropicDispersionSimplestForm}
\epsilon_{\mu}(\vec{p}) = a_{1\mu}p^3 + a_{2\mu}p_x \,,
\end{equation}
where $\mu$ is the band index, $a_{1\mu}$ and $a_{2\mu}$ are dispersion coefficients, $p \equiv |\vec{p}| = (p_x^2+p_y^2+p_z^2)^{\frac{1}{2}}$. A number of materials for the quadratic and cubic Weyl/Dirac semimetals have been recently proposed and extensively studied. Examples of quadratic Weyl/Dirac semimetals include HgCr$_2$Se$_4$, SiSr$_2$, band-inverted $\alpha-$Sn and PdSb$_2$ \cite{HigherOrder_1, HigherOrder_4, HigherOrder_6}, and examples of cubic Weyl/Dirac semimetals include LiO$_\text{S}$O$_3$ and quasi-one-dimensional molybdenum monochalcogenide compounds A$^\text{I}$(MoX$^{\text{VI}}$)$_3$, where A$^\text{I}$ = Na, K, Rb, In or Tl, X$^{\text{VI}}$ = S, Se or Te \cite{HigherOrder_7, HigherOrder_9, HigherOrder_12, HigherOrder_13}. Moreover, various novel quantum phenomena are predicted to take place in these semimetals, such as charge density wave, non-Fermi liquid, and topological superconductivity \cite{HigherOrder_2, HigherOrder_8, HigherOrder_9, HigherOrder_12, HigherOrder_13, HigherOrder_14, 1D_1, 1D_2, 1D_3, 1D_4, 1D_5, 1D_6, 1D_7, 1D_8, 1D_9, 1D_10}. It is therefore timely and of theoretical importance to study Kondo problems in these higher-order fermion (KHOF)
systems, in particular, using analytically exact methods.
 
In ideal KHOF systems, the Fermi energy is located at the Dirac point (DP), with higher-order dispersions along certain directions in momentum space. In these cases, the aforementioned linearization approximation in AL's original CFT approach becomes inapplicable. On one hand, it generates an artificial flat bands along certain directions, as shown by Fig.~\ref{fig:MappingToComplexPlane}(b). Clearly, the flat band is not sufficient  to capture the realistic low-energy excitations of the higher-order fermions. On the other hand, in contrast to the Kondo problems in normal metals with a single band, anisotropic multi-bands with touching nodes must be taken into account. These unique features of the KHOF systems pose severe challenges for the conventional CFT approach. \textit{Does there exist any alternative CFT scheme that overcome these difficulties?}

In this work, we propose a full-energy mapping CFT (FEMCFT) method. This method solves all the above problems, and produces exact solutions to Kondo problems in a large class of exotic semimetals, in particular, the KHOF models. In contrast with the conventional CFT scheme that only establishes conformal invariance for the linearized low-lying excitations within some artificially-introduced energy cut-off $\Lambda$ near the Fermi surface, our approach is able to map the full energy spectrum of the KHOF model into a form that observes conformal invariance. Consequently, instead of artificially introducing a cut-off $\Lambda$, and only mapping the low energy degrees of freedom of the bath bounded by $\Lambda$ to the complex plane, we can map the full energy spectrum into the complex plane without introducing any artificial cut-offs in the spectrum, as shown by Fig.~\ref{fig:MappingToComplexPlane}(c) and \ref{fig:MappingToComplexPlane}(d). As a result, our approach is free from the flat band issue shown in Fig.~\ref{fig:MappingToComplexPlane}(b). An additional advantage of our method is that, by rigorously taking into account the entire energy spectrum of the bath, we can make more accurate predictions about the low-energy fixed point and the thermodynamic quantities at finite temperatures, compared to conventional CFT techniques.

We emphasize that our FEMCFT approach is analytically exact for ideal isotropic KHOF systems. It also works well for highly realistic anisotropic KHOF materials, after systematically treating the anisotropic effects as corrections to the isotropic parts. Our work therefore rigorously solves the KHOF and related models for higher-order topological semimetals, and significantly advances the scope of CFT approaches to many-body resonances, bringing about new applications in previously inaccessible systems.

The remaining part of the manuscript is organized as follows. In Sec.II, we define the general Hamiltonian for the KHOF systems, which describes a multichannel Kondo impurity in higher-order Weyl/Dirac systems. In Sec.III, we present our FEMCFT approach in the isotropic limit, which establishes an exact mapping of an ideal isotropic KHOF system to a 2D CFT in the complex plane. The calculations of thermodynamic quantities are discussed in Sec.IV. In Sec.V, we tackle realistic anisotropic KHOF systems, and demonstrate our method with an explicit example, namely, the Kondo problem in the anisotropic cubic Weyl/Dirac fermion system in three dimensions, whose dispersion is given by (\ref{AnisotropicDispersionSimplestForm}), which has attracted great interests recently. 

\section{General Hamiltonian for KHOF systems}
For the sake of generality, we consider a  KHOF system in $n$ spatial dimensions, whose Hamiltonian is given by
\begin{equation}\label{H=H0+HI}
H = H_0 + H_I \,,
\end{equation}
where $H_0$ is the bath Hamiltonian, and $H_I$ is the interaction Hamiltonian. $H_0$ is given by
\begin{equation}\label{eq:StartingPointH0}
H_0 = \int d^n \vec{p} \, c^{\dagger}_{\alpha, i, \mu}(\vec{p}) c_{\alpha, i, \mu}(\vec{p}) \epsilon_{\mu}(\vec{p}) \,,
\end{equation}
where $\vec{p} = (p_1,...,p_n)$, $c^{\dagger}_{\alpha, i, \mu}(\vec{p})$ and $c_{\alpha, i, \mu}(\vec{p})$ are conduction electron creation and annihilation operators respectively, with $\alpha \in \{\uparrow, \downarrow \}$ labelling the up/down spin index, $i \in \{1,2,...,k \}$  labelling the channel index, and $\mu \in \{+, -\}$ labelling the band index. Summation over repeated indices is implied.

We consider a general higher-order dispersion $\epsilon_{\mu}(\vec{p})$ of the form
\begin{equation}\label{3DSampleDispersionRelation_Sec2}
\epsilon_{\mu}(\vec{p}) = a_{1\mu}p^n + a_{2\mu}p_1 \,,
\end{equation}
where $p \equiv |\vec{p}|$. The constants $a_{1\mu}$ and $a_{2\mu}$ satisfy $a_{1+} = -a_{1-}$, $a_{2+} = -a_{2-}$. The dispersion (\ref{3DSampleDispersionRelation_Sec2}) is of particular interest, because for $n=3$, it reduces to a cubic fermion system in three-dimensions (\ref{AnisotropicDispersionSimplestForm}). For $n=2$, it produces a quadratic band crossing point, which also has attracted great interests in the past decade \cite{QBCP_1, QBCP_2, QBCP_3}. Moreover, for $n \geq 3$, it describes even higher order emergent fermions which can be realized in artificial electric circuits with auxiliary dimensions \cite{ElectricCircuit}.

$H_I$ describes the exchange interaction between an impurity spin and the bath electrons:
\begin{equation}\label{eq:StartingPointHI}
H_I = \frac{\lambda_i \vec{S}}{(2\pi)^n} \cdot \sum_{\mu, \nu} \int d^n \vec{p} \, d^n \vec{p}' c^{\dagger}_{\alpha, i, \mu}(\vec{p}) \frac{\vec{\sigma}_{\alpha\beta}}{2} c_{\beta, i, \nu}(\vec{p}') \,,
\end{equation}
where $\vec{S}$ is the spin of the magnetic impurity, $\lambda_i$ is the Kondo coupling constant for the $i$-th channel.

Eq. (\ref{H=H0+HI}) - (\ref{eq:StartingPointHI}) constitute the most general KHOF model in $n$-dimensions. Our focus is to look for exact CFT solutions to Eq. (\ref{H=H0+HI}) - (\ref{eq:StartingPointHI}).

\section{FEMCFT approach in the isotropic limit}
In this section, we outline our FEMCFT approach for an ideal isotropic KHOF system, i.e. the case where $a_{2\mu} = 0$ in (\ref{3DSampleDispersionRelation_Sec2}), so that
\begin{align}\label{DispersionRelation_n=m}
\epsilon_{\mu}(p) = a_{1\mu}p^n\,.
\end{align}
More details of the derivations in this section can be found in Appendix~\ref{Appdendix:IsotropicDetailedDerivations}. We shall treat the anisotropic case $a_{2\mu} \neq 0$ in Sec.V. 

We expand $c_{\alpha, i, \mu}(\vec{p})$ as a linear combination of the spherical harmonics, which form a complete set of orthonormal functions on the $(n-1)-$ sphere $S^{n-1}$:
\begin{align}\label{VecInTermsOfMagnitude}
& c_{\alpha, i, \mu}(\vec{p}) = \frac{1}{p^{\frac{n-1}{2}}} \cdot \nonumber\\
& \sum_{l_1,...,l_{n-1}} Y_{l_1,...,l_{n-1}}(\theta_1,...,\theta_{n-1}) c_{l_1, ..., l_{n-1}, \alpha, i, \mu}(p) \,.
\end{align}
Here $Y_{l_1,...,l_{n-1}}(\theta_1,...,\theta_{n-1})$ are the spherical harmonics in $n$ dimensions, see Appendix~\ref{Appdendix:nDimSphericalHarmonics} and \cite{nDimSphericalHarmonics} for detailed derivations of their important properties relevant for this work. $\theta_1,...,\theta_{n-1}$ are the $n-1$ angular coordinates of $S^{n-1}$, with $0 \leq \theta_1 < 2\pi$, $0 \leq \theta_j \leq \pi$ for $j=2,...,n-1$. The integers $|l_1| \leq l_2 \leq ... \leq l_{n-1}$ denote different partial waves, and are the analogues of the quantum numbers $m, l$ in the three-dimensional spherical harmonics $Y_l^m (\theta, \phi)$: in 3 dimensions, $n=3$, $l_1 \equiv m, l_2 \equiv l$, $\theta_1 \equiv \phi$ is the azimuthal angle, and $\theta_2 \equiv \theta$ is the polar angle. $H_0$ and $H_I$ can be written in terms of $c_{l_1,...,l_{n-1},\alpha,i,\mu}(p)$ as
\begin{align}\label{H_IBranchPt}
H_{I} = \frac{\lambda_i \vec{S}}{\Gamma(\frac{n}{2})2^n\pi^{\frac{n}{2}}} \cdot \sum_{\mu, \nu} \int & dpdp' p^{\frac{n-1}{2}} p'^{\frac{n-1}{2}} \nonumber\\
& c^{\dagger}_{0,...,0,\alpha,i,\mu}(p) \vec{\sigma}_{\alpha\beta} c_{0,...,0,\beta,i,\nu}(p')\,,
\end{align}
\begin{align}\label{H_0BranchPt}
H_0 = \int dp \, c^{\dagger}_{0, ..., 0, \alpha, i, \mu}(p) c_{0, ..., 0, \alpha, i, \mu}(p) \epsilon_{\mu} (p) \,.
\end{align}

Next we perform a change of variables on the fields $c_{l_1,...,l_{n-1},\alpha,i,\mu}(p)$ from $p$ to $\epsilon$, by defining
\begin{equation}\label{definition_epsilonfield}
c_{l_1, ..., l_{n-1}, \alpha, i, \mu}(\epsilon_{\mu}) \equiv \left| \frac{d\epsilon_{\mu}(p)}{dp} \right|^{-\frac{1}{2}} c_{l_1, ..., l_{n-1}, \alpha, i, \mu}(p) \,.
\end{equation}
In this way, the $c_{l_1, ..., l_{n-1}, \alpha, i, \mu}(\epsilon_{\mu})$ fields also satisfy the proper fermionic anti-commutation relations. With the definition
\begin{align}
\Phi_{0,...,0,\alpha,i}(\epsilon) \equiv
\begin{cases}
c_{0, ..., 0, \alpha, i, +}(\epsilon) & \text{if } \epsilon \geq 0 \\
c_{0, ..., 0, \alpha, i, -}(\epsilon) & \text{if } \epsilon < 0
\end{cases} \,,
\end{align}
we combine the fields from the + and - bands into one single composite fermionic field $\Phi_{0,...,0,\alpha,i}(\epsilon)$. In terms of $\Phi_{0,...,0,\alpha,i}(\epsilon)$, $H_0$ and $H_I$ become
\begin{align}
H_0 = \int_{-\infty}^{\infty} d\epsilon \, \Phi^{\dagger}_{0, ..., 0, \alpha, i}(\epsilon) \Phi_{0, ..., 0, \alpha, i}(\epsilon) \epsilon \,,
\end{align}
\begin{align}
H_I & = \frac{\lambda_i \vec{S}}{an\Gamma(\frac{n}{2})2^n\pi^{\frac{n}{2}}} \nonumber\\
& \cdot \int_{-\infty}^{\infty} d\epsilon \int_{-\infty}^{\infty} d\epsilon' \Phi^{\dagger}_{0,...,0,\alpha,i}(\epsilon) \vec{\sigma}_{\alpha\beta} \Phi_{0,...,0,\beta,i}(\epsilon') \,,
\end{align}
where $a \equiv |a_{1\mu}|$. Notice that we have combined the $+$ and $-$ bands into a single band, and eliminated the band index $\mu \in \{+, -\}$ from our model.

We then define the left and right moving fields, respectively, as
\begin{align}\label{Definition_of_r_Fields}
&\Psi_{\leftarrow \alpha i} (r) \equiv \int_{-\infty}^{\infty} d \epsilon \, e^{-i \epsilon r} \Phi_{0, ..., 0, \alpha, i}(\epsilon) \nonumber\\
&\Psi_{\rightarrow \alpha i} (r) \equiv \int_{-\infty}^{\infty} d \epsilon \, e^{+i \epsilon r} \Phi_{0, ..., 0, \alpha, i}(\epsilon) \,.
\end{align}
Also, by introducing the imaginary time $\tau \equiv it$, we define the complex plane $\mathbb{C}$, to which we map our KHOF model, by
\begin{equation}\label{DefinitionOfComplexPlane}
\mathbb{C} = \{ z \equiv \tau + ir \} \,.
\end{equation}
i.e. The horizontal axis of $\mathbb{C}$ is the imaginary time $\tau$, and the vertical axis of $\mathbb{C}$ is $r$ introduced in (\ref{Definition_of_r_Fields}). We then view $\Psi_{\leftarrow/\rightarrow \alpha i} (r)$ in the Heisenberg picture, so that they now have time-dependence, and live on $\mathbb{C}$. $\Psi_{\leftarrow \alpha i} (\tau, r)$ and $\Psi_{\rightarrow \alpha i} (\tau, r)$ are related to each other by
\begin{equation}
\Psi_{\rightarrow \alpha i}(\tau, r) = \Psi_{\leftarrow \alpha i}(\tau, -r) \,,
\end{equation}
so $\Psi_{\rightarrow \alpha i}(\tau, r)$ can be eliminated in terms of $\Psi_{\leftarrow \alpha i} (\tau, r)$. $H_0$ and $H_I$ can be written in terms as $\Psi_{\leftarrow \alpha i} (\tau, r)$ as
\begin{align}\label{H0_Final}
H_0 = \frac{1}{2\pi} \int^{\infty}_{-\infty} dr \left( \Psi^{\dagger}_{\leftarrow \alpha i}(\tau, r) i\frac{\partial}{\partial r} \Psi_{\leftarrow \alpha i}(\tau, r) \right) \,,
\end{align}
\begin{align}\label{HI_Final}
H_{I} = & \frac{\lambda_i \vec{S}}{an\Gamma(\frac{n}{2})2^n\pi^{\frac{n}{2}}} \cdot \Psi^{\dagger}_{\leftarrow \alpha i}(\tau, 0) \vec{\sigma}_{\alpha\beta} \Psi_{\leftarrow \beta i}(\tau,0) \,.
\end{align}
The $n$-dimensional KHOF model is thus mapped into the complex plane $\mathbb{C}$ defined in (\ref{DefinitionOfComplexPlane}), with $H_0$ and $H_I$ taking the forms of (\ref{H0_Final}) and (\ref{HI_Final}) respectively. These are in similar forms as the Hamiltonians mapped from single band normal metals. In particular, we see from (\ref{HI_Final}) that the Kondo exchange coupling remains short-ranged in the new fermionic degrees of freedom; the impurity spin $\vec{S}$ only couples to the new fermionic field $\Psi_{\leftarrow \alpha i}(\tau,r)$ at $r=0$. This means that $H_I$ is only confined to the boundary $r=0$, with $H=H_0$ in the bulk $r \neq 0$, and the problem is suitable for further analysis by techniques in 2D boundary CFT.

We note that our FEMCFT approach is analytically exact; we have not introduced any artificial cut-off $\Lambda$ in the momentum or energy. Our integrals $\int_0^{\infty} dp$ and $\int_{-\infty}^{\infty} d\epsilon$ are over the entire spectrum. Also, we have transformed the 2-band KHOF system into an effective one-band system. This further facilitates the analysis of KHOF systems using CFT techniques.

\section{Thermodynamic Quantities}
After mapping KHOF systems into the form (\ref{H0_Final}) + (\ref{HI_Final}) on the complex plane via our FEMCFT approach, we can proceed to determine the thermodynamic quantities at $T>0$. This is carried out via AL's standard CFT techniques, which we briefly summarize here. More details on their approach can be found in \cite{AL1,AL2,AL3,AL4,AL5,AL6,AL7,VonDelft}. The underlying geometry of  the $T>0$ physics is the infinite cylinder with circumference $\beta = 1/T$. The system is further mapped from the complex plane onto the infinite cylinder via a conformal map. The complete list of boundary operators for the system with a particular boundary condition can be then obtained by applying ``double fusion" to the free system (i.e. the system with trivial boundary condition). From the list of boundary operators, we can determine the leading irrelevant operator with coupling constant $\tilde{\lambda}$, whose Green's functions allow us to compute the thermodynamic quantities of interest. The circumference of the cylinder $\beta = 1/T$ enters the calculation of the thermodynamic quantities as a finite size of the system, giving rise to their $T$-dependences. For example, the resistivity $\rho$ is found as $\rho = \rho_U \left\{ 1- \alpha \tilde{\lambda}^2T^2 + ...\right\}$ in the Fermi liquid case and $\rho = \rho_U\frac{1-S^{(1)}}{2} \left\{ 1+ \alpha \tilde{\lambda} T^{\frac{2}{2+k}} + ... \right\}$ in the non-Fermi liquid case. Here $k$ is the number of channels, $S^{(1)} = \frac{\cos\{\pi(2s+1)/(2+k)\}}{\cos\{\pi/(2+k)\}}$ with $s$ being the impurity spin, $\rho_U$ is the unitary limit resistivity, i.e. the greatest resistivity possibly achievable, and $\alpha$ is a dimensionless constant, with $\alpha = 4\sqrt{\pi}$ in the special case $k =2$. These results are in excellent agreements with results obtained from numerical renormalization group (NRG) analysis.

We remark that, although our $T>0$ thermodynamic quantities take the same form as AL's, they are valid over greater ranges of temperatures compared to AL's. This is because AL's original approach to Kondo problems in normal metals requires the introduction of a narrow cutoff $\Lambda$ about the Fermi surface. Thus in their calculated thermodynamic quantities, AL can only consider temperatures $T$ satisfying $T \ll \Lambda v_F$, and ignore terms of order $T/\Lambda v_F$. In comparison, we did not introduce any artificial cut-offs in the spectrum. As a result, our calculated thermodynamic quantities are more accurate and are valid over greater ranges of temperatures, without the restriction of $T \ll \Lambda v_F$. Moreover, conventional CFT approaches are not applicable at all in KHOF systems with coinciding Fermi energies and DPs. Our theory fills this research gap, and enables an exact CFT analysis for these novel phases with emergent higher-order fermions.

\section{Realistic cases with anisotropy}
Realistic topological semimetals with higher order fermions are always anisotropic.  In this section, we therefore present a general framework to treat this anisotropy, followed by an explicit application of our framework to a concrete example, namely the anisotropic cubic fermion model in three dimensions, realized by setting $n=3$ and $a_{2\mu}\neq0$ in (\ref{3DSampleDispersionRelation_Sec2}). More details of the derivations can be found in Appendix~\ref{Appendix:Anisotropic}.

For anisotropic KHOF systems, $H_I$ remains the same as before, since the dispersion relation does not enter the expression of $H_I$. However, $H_0$ can no longer be reduced to the simple form (\ref{H_0BranchPt}), due to the lack of spherical symmetry. In general, we can express $H_0$ into a matrix form in the partial wave basis, i.e.,
\begin{equation}\label{H0MatrixForm}
H_0 = \int dp \, C^{\dagger}_{\alpha , i, \mu}(p) A_{\mu}(p) C_{\alpha , i, \mu}(p), \
\end{equation}
where $C_{\alpha , i, \mu}(p)$ is a column vector whose $L-$th element is given by $c_{L, \alpha , i, \mu}(p)$, where for brevity we have used the multi-index notation $L \equiv (l_1,...,l_{n-1})$. $C^{\dagger}_{\alpha , i, \mu}(p)$ is the Hermitian conjugate of $C_{\alpha , i, \mu}(p)$ and $A_{\mu}(p)$ is a matrix whose $(L, L')-$th element, $A_{L,L',\mu}(p)$, describes the coupling between the $L$ and $L'-$th partial waves. We remark that for all Hermitian systems, the only partial waves that couple to the impurity are those with $l_1=0$. Thus the multi-index $L$ only includes $(l_1= 0, l_2,...,l_{n-1})$. In particular, in three dimensions, $L$ only includes $(m=0, l) \rightarrow l$, i.e. the multi-index $L$ reduces to the single index $l$: $L=l$.

\begin{figure}
\includegraphics[width=\linewidth]{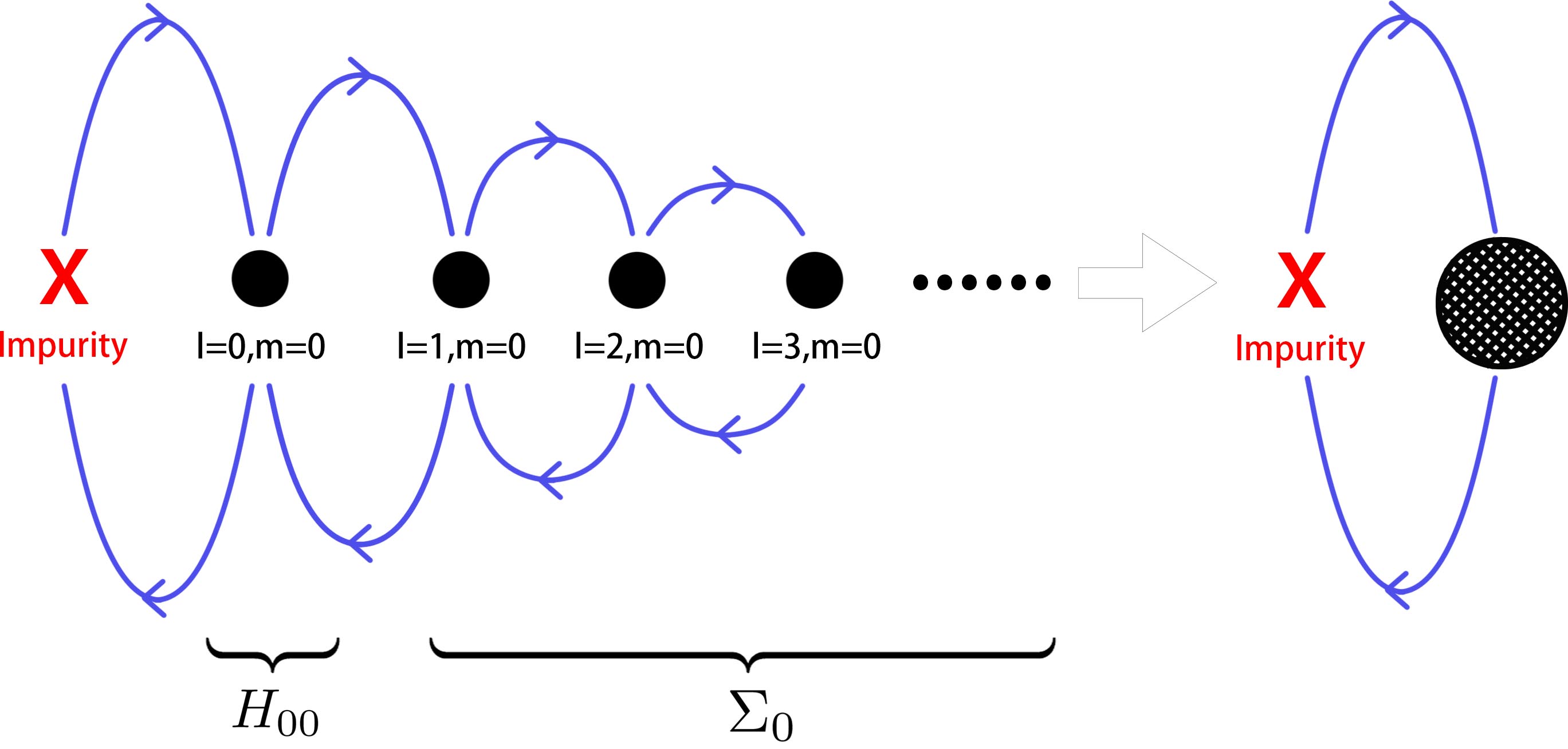}
\caption{\label{fig:Hopping} The couplings between the impurity and the different partial waves in the three-dimensional KHOF system with dispersion Eq.\eqref{3DSampleDispersionRelation} is illustrated by a hierarchical one-dimensional chain. Only the $m=0$ partial waves are relevant, which is a property universal for all Hermitian systems. The impurity only couples directly to the $l=0$, $m=0$ partial wave, which further couples to higher-order $l$-components via the nearest ``hoppings". The hopping coefficient decays with $l$.  Clearly, the anisotropy is manifested by the higher-order corrections to the isotropic component. By integrating out the $l\neq0$ components, the impurity is effectively coupled to the renormalized $l=0$ $m=0$ component, where the anisotropic effects have been absorbed as corrections to the isotropic part. }
\end{figure}

We separate $H_0$ in (\ref{H0MatrixForm}) into two terms
\begin{equation}
H_0 = H_{00} + \Sigma_0 \,,
\end{equation}
where the term $H_{00}$ consists of only the $L = (0,...,0)$ partial waves, thus describing the ``isotropic part" of $H_0$.
 $\Sigma_0$ consists of all the remaining partial waves with $L\neq(0,...,0)$, and captures the ``anisotropic part" of $H_0$, as well as its couplings to $H_{00}$, as shown by left picture of Fig.2.

Due to anisotropy, partial waves with different quantum numbers are coupled to each other, forming a hierarchical structure depicted in the left picture of Fig.~\ref{fig:Hopping} (in general, different models can give rise to different coupling hierarchical structures. The left picture of Fig.~\ref{fig:Hopping} shows the case for the cubic fermion model in three dimensions (\ref{3DSampleDispersionRelation}), which suffices to illustrate our framework). As shown in Fig.~\ref{fig:Hopping}, the impurity firstly couples to the isotropic sector, which further couples to the higher components. Owing to the hierarchical nature of the couplings, we can integrate out the anisotropic part by down-folding the Hamiltonian matrix in Eq.\eqref{H0MatrixForm} to the isotropic subspace. It can be shown that in the static limit, the anisotropic effects of $\Sigma_0$ can be casted into an additional renormalization term $\epsilon_{\Sigma \mu} (p)$, which serves as a modification to the isotropic part of the dispersion relation $\epsilon_{\mu}(p)$.

To further enable a CFT analysis, we approximate the value of $\epsilon_{\Sigma \mu}(p)$ by evaluating it at the Fermi wave number, $p_F$. This fully captures the low-energy Kondo physics, since only the excitations near the Fermi point are important at low temperatures. It should be noted that, although this approximation is in the same spirit as AL's linearization approximation near the Fermi surface, it is made only to the anisotropic part of the dispersion. Our method still admits an exact treatment of the isotropic part by mapping the  entire spectrum into a two-dimensional CFT. Therefore, it is free from the flat band issue illustrated in Fig.~\ref{fig:MappingToComplexPlane}(b), and overcomes the major difficulty present in conventional CFT techniques.

The above procedures generate an renormalized bath coupled to the impurity, as indicated by the right picture of Fig.~\ref{fig:Hopping}. The renormalized bath enjoys the effective dispersion relation $\tilde{\epsilon}_{\mu}(p) \equiv \epsilon_{\mu}(p) + \epsilon_{\Sigma \mu} (p_F)$, which contains the isotropic part of the dispersion, as well as the corrections from the anisotropy. Correspondingly, the bath is now effectively described by
\begin{align}
& H_0  = H_{00} + \Sigma_0 \approx \int dp \, c^{\dagger}_{0, ..., 0, \alpha, i, \mu}(p) c_{0, ..., 0, \alpha, i, \mu}(p) \tilde{\epsilon}_{\mu}(p)\,.
\end{align}
The above steps constitute a general framework that takes into account the anisotropy in the CFT analysis.

We now apply our approach to a concrete example, namely the  cubic fermion system in three dimensions. This is realized by setting $n=3$ and turning on $a_{2\mu}$ in (\ref{3DSampleDispersionRelation_Sec2}), namely
\begin{equation}\label{3DSampleDispersionRelation}
\epsilon(\vec{p}) = a_{1\mu}p^3 + a_{2\mu}p_1 \,,
\end{equation}
which is linear in the $p_1-$direction but cubic in the orthogonal plane, a typical low-energy dispersion of cubic fermion materials.

This model admits a simple coupling hierarchical structure as shown in Fig.~\ref{fig:Hopping}. Here, the impurity is coupled only to the isotropic component, which is further connected to the higher-order partial waves via nearest-neighbour hoppings. Moreover, it can be shown that the hopping strength decays as $l$ increases.  As a result,  $A_{\mu}(p)$ in (\ref{H0MatrixForm}) can be casted into the simple tridiagonal form
\begin{align}\label{TridiagonalMatrix}
& A_{\mu}(p) =
  \left(\begin{array}{@{}cccccc@{}}
    a_{1\mu}p^3 & \frac{a_{2\mu}p}{\sqrt{3}} & & & & \\
    \frac{a_{2\mu}p}{\sqrt{3}} & a_{1\mu}p^3 & \frac{2a_{2\mu}p}{\sqrt{15}} & & & \\
     & \frac{2a_{2\mu}p}{\sqrt{15}} & a_{1\mu}p^3 & \frac{3a_{2\mu}p}{\sqrt{35}} & & \\
     & & \frac{3a_{2\mu}p}{\sqrt{35}} & a_{1\mu}p^3 & \frac{4a_{2\mu}p}{\sqrt{63}} & \\
     & & & \ddots & \ddots & \ddots
  \end{array}\right) \,.
\end{align}

We then integrate out the higher-order partial waves by down-folding the matrix $A_{\mu}(p)$ to the isotropic subspace. Interestingly, because each entry of $A_{\mu}(p)$  is proportional to $p$, the renormalization $\epsilon_{\Sigma_{\mu}}(p)$ is also found to be proportional to $p$, and thus vanishes at the DP with $p_F=0$.  This essentially indicates that the contributions from different high-order partial waves display a ``destructive interference", in the sense that they completely cancel with each other at the DP. This is an interesting feature of cubic Dirac fermions, which greatly simplifies the treatment of anisotropy.

The effective dispersion relation is eventually obtained as
\begin{equation}\label{FinalEffectiveDispersionRelation}
\tilde{\epsilon}_{\mu}(p) = a_{1\mu}p^3\,,
\end{equation}
which is clearly of the form (\ref{DispersionRelation_n=m}) in three dimensions. As a result, all subsequent derivations following Eq. (\ref{DispersionRelation_n=m}) hold, and our FEMCFT approach for isotropic KHOF systems  applies. $H_0$ and $H_I$ will be mapped to the forms of (\ref{H0_Final}) and (\ref{HI_Final}) respectively. Correspondingly, the impurity ground state and the thermodynamic quantities  can be readily obtained using the previously discussed methods.

\section{Conclusion}
In conclusion, we have developed a full-energy mapping conformal field theory (FEMCFT) approach to tackle Kondo problems in higher order fermion and related systems. Our FEMCFT approach is capable of solving these models exactly, which are inaccessible by using conventional CFT approaches. Moreover, the FEMCFT approach is able to make more accurate predictions about the thermodynamic quantities over greater ranges of temperatures. We applied our FEMCFT approach to a specific KHOF system, namely the cubic Weyl/Dirac fermion system in three dimensions. The high efficiency of our approach to realistic KHOF systems is clearly justified.

The FEMCFT approach significantly broadens the scope of existing CFT methods in the study of Kondo problems. We anticipate developments of novel CFT techniques for treating Kondo problems in more complicated systems, such as those displaying pseudogaps, as possible future directions of research. Such advancements would undoubtedly further fortify the connections between the mathematically elegant CFTs and the physically intriguing fixed points in strong-correlated problems, such as novel quantum criticalities and many-body resonances in strong-coupling limits.
\begin{acknowledgments}
We are grateful to W. Su, Bin Tai, Feng Tang, Tigran Sedrakyan, and Y. X. Zhao for fruitful discussions. This work was supported by the Jiangsu Postdoctoral Research Grant (Grant No. 2020Z019), the Youth Program of National Natural Science Foundation of China (No. 11904225) and the National Key R\&D Program of China (Grant No. 2017YFA0303200).
\end{acknowledgments}

\appendix
\section{Spherical Harmonics in Higher Dimensions}\label{Appdendix:nDimSphericalHarmonics}
We review the properties of the $n$-dimensional spherical harmonics used in the main section. Some of this material can also be found in \cite{nDimSphericalHarmonics}.

The $n$-dimensional spherical harmonics are the eigenfunctions of $\Delta_{S^{n-1}}$, the Laplace-Beltrami operator on the sphere $S^{n-1}$, which is the angular part of the $n$-dimensional Laplacian operator. The Laplace-Beltrami operator can be defined recursively:
\begin{align}
\Delta_{S^{j}} = & \left( \frac{1}{\sin^{j-1}\theta_{j}} \right) \frac{\partial}{\partial \theta_{j}} \left( \sin^{j-1} \theta_{j} \frac{\partial}{\partial \theta_{j}} \right)
 + \frac{\Delta_{S^{j-1}}}{\sin^{2} \theta_{j}}
\end{align}
for $2 \leq j \leq n-1$, and
\begin{equation}\label{LB1}
\Delta_{S^1} = \frac{\partial^2}{\partial \theta_1^{\, 2}} \,,
\end{equation}
where $\theta_1,...,\theta_{n-1}$ are the $n-1$ angular coordinates in the $n$-dimensional spherical coordinates, with $0 \leq \theta_1 < 2\pi$, $0 \leq \theta_2,...,\theta_{n-1} \leq \pi$. An $n$-dimensional spherical harmonic $Y(\theta_1,..., \theta_{n-1})$ of polynomial degree $l_{n-1}$, where $l_{n-1}$ is a non-negative integer, has eigenvalue $-l_{n-1}(l_{n-1}+n-2)$:
\begin{align}\label{MainDiffyQ}
\Delta_{S^{n-1}} \, Y(\theta_1,..., \theta_{n-1}) = & -l_{n-1}(l_{n-1}+n-2) \nonumber\\
& Y(\theta_1,..., \theta_{n-1}) \,.
\end{align}
This equation can be solved by separation of variables. By writing
\begin{equation}
Y(\theta_1,..., \theta_{n-1}) = Y'(\theta_1,..., \theta_{n-2})\Theta_{n-1} (\theta_{n-1}) \,,
\end{equation}
where $Y'(\theta_1,..., \theta_{n-2})$ is a spherical harmonic of degree $l_{n-2}$ on $S^{n-2}$, and $\Theta_{n-1} (\theta_{n-1})$ is a function that only depends on $\theta_{n-1}$, to be further determined. Because $Y'(\theta_1,..., \theta_{n-2})$ is a factor of $Y(\theta_1,..., \theta_{n-1})$, its polynomial degree must not be larger than that of $Y(\theta_1,..., \theta_{n-1})$, thus $l_{n-2} \leq l_{n-1}$. We can now write (\ref{MainDiffyQ}) as
\begin{align}\label{SepOfVar_OverallEqn}
& \frac{\sin^{4-n} \theta_{n-1}}{\Theta_{n-1}} \frac{\partial}{\partial \theta_{n-1}} \left( \sin^{n-2} \theta_{n-1} \frac{\partial \Theta_{n-1}}{\partial \theta_{n-1}} \right) \nonumber\\
+ & l_{n-1}(l_{n-1}+n-2)(\sin^2 \theta_{n-1}) + \frac{1}{Y'} \Delta_{S^{n-2}}Y' = 0 \,.
\end{align}
The first two terms only depend on $\theta_{n-1}$, and the third term only depend on $\theta_1,..., \theta_{n-2}$. Since $Y'(\theta_1,..., \theta_{n-2})$ is a spherical harmonic of degree $l_{n-2}$ on $S^{n-2}$, we have
\begin{align}\label{SepOfVar_2ndEqn}
\Delta_{S^{n-2}} \, Y'(\theta_1,...,\theta_{n-2}) = & -l_{n-2}(l_{n-2}+n-3) \cdot \nonumber\\
& Y'(\theta_1,...,\theta_{n-2}) \,,
\end{align}
so the third term in (\ref{SepOfVar_OverallEqn}) must equal to $-l_{n-2}(l_{n-2}+n-3)$. As a result, the first two terms of (\ref{SepOfVar_OverallEqn}) must equal to $+l_{n-2}(l_{n-2}+n-3)$:
\begin{align}\label{SepOfVar_1stEqn}
& 0 = \sin^{4-n} \theta_{n-1} \frac{\partial}{\partial \theta_{n-1}} \left( \sin^{n-2} \theta_{n-1} \frac{\partial \Theta_{n-1}}{\partial \theta_{n-1}} \right) + \Theta_{n-1} \cdot \nonumber\\
& \left\{ l_{n-1}(l_{n-1}+n-2)(\sin^2 \theta_{n-1})
- l_{n-2}(l_{n-2}+n-3) \right\} \,.
\end{align}
Now define
\begin{align}
{}_{b}\bar{P}_{c}^{a} (\theta) \equiv & \sqrt{\frac{(2c+b-1)(a+b+c-2)!}{2(c-a)!}} \nonumber\\
& \sin^{\frac{2-b}{2}}(\theta) P^{-(a+ \frac{b-2}{2})}_{c+ \frac{b-2}{2}}(\cos \theta) \,,
\end{align}
where $P^{m}_{l}(x)$ is the associated Legendre function of the first kind. In our case $m, l$ are generalized to take on integer or half-integer values. By noting that $P^{m}_{l}(x)$ solves the Legendre equation:
\begin{align}
& (1-x^2) \frac{d^2}{dx^2} P^{m}_{l}(x) -2x \frac{d}{dx} P^{m}_{l}(x) \nonumber\\
+ & \left\{ l(l+1) - \frac{m^2}{1-x^2} \right\} P^{m}_{l}(x) = 0 \,,
\end{align}
the solution to equation (\ref{SepOfVar_1stEqn}) is
\begin{equation}
\Theta_{n-1} = {}_{n-1}\bar{P}_{l_{n-1}}^{l_{n-2}} (\theta_{n-1}) \,.
\end{equation}
Equation (\ref{SepOfVar_2ndEqn}) is in the same form as (\ref{MainDiffyQ}) and can be solved recursively by repeating the procedure up to now, yielding $Y'(\theta_1,...,\theta_{n-2}) = Y''(\theta_1,...,\theta_{n-3})\Theta_{n-2}(\theta_{n-2})$ and $\Theta_{n-2}(\theta_{n-2}) = {}_{n-2}\bar{P}_{l_{n-2}}^{l_{n-3}} (\theta_{n-2})$. Eventually we obtain $\Theta_{2}(\theta_{2}) = {}_{2}\bar{P}_{l_{2}}^{l_{1}} (\theta_{2})$, and the equation
\begin{align}
\Delta_{S^1} \, \Theta_1(\theta_1) = -l_1^2 \, \Theta_1(\theta_1)
\end{align}
where $\Delta_{S^1}$ is given by (\ref{LB1}). Its solution is
\begin{equation}
\Theta_1 (\theta_1) = e^{il_1 \theta_1} \,.
\end{equation}
Thus we arrive at
\begin{align}
Y_{l_1,...,l_{n-1}} (\theta_1,...,\theta_{n-1}) = & A \prod^{n-1}_{j=1} \Theta_j(\theta_j) \nonumber\\
= & A e^{il_1 \theta_1}  \prod^{n-1}_{j=2} {}_{j}\bar{P}_{l_j}^{l_{j-1}} (\theta_j) \,,
\end{align}
where $|l_1| \leq l_2 \leq ... \leq l_{n-1}$. The integers $l_1, l_2, ... l_{n-1}$ are analogues of the quantum numbers $m, l$ in the 3-dimensional spherical harmonics $Y_l^m (\theta, \phi)$. Indeed, $|l_1| \leq l_2$ is due to the same reason as $|m| \leq l$ in the 3D case. In particular, in 3D, $n=3$, $l_1 \equiv m, l_2 \equiv l, \theta_1 \equiv \phi$ is the azimuthal angle, and $\theta_2 \equiv \theta$ is the polar angle. Also, $A$ is a normalization constant to be determined. Before determining $A$, let us show that the $n$-dimensional spherical harmonics satisfy the orthonormality condition:
\begin{align}\label{Orthogonality_Appendix}
& \int Y^*_{l_1,...,l_{n-1}} (\theta_1,...,\theta_{n-1}) Y_{l'_1,...,l'_{n-1}} (\theta_1,...,\theta_{n-1}) d\Omega_{n-1} \nonumber\\
= & \delta_{l_1l'_1}\delta_{l_2l'_2}...\delta_{l_{n-1}l'_{n-1}} \,.
\end{align}
The normality condition in (\ref{Orthogonality_Appendix}) can help us in determining the normalization constant $A$. Consider the integral
\begin{align}\label{Orthogonality_Appendix_MasterEqn}
& \int Y^*_{l_1,...,l_{n-1}} (\theta_1,...,\theta_{n-1}) Y_{l'_1,...,l'_{n-1}} (\theta_1,...,\theta_{n-1}) d\Omega_{n-1} \nonumber\\
= & |A|^2 \int e^{-i l_1 \theta_1}e^{i l'_1 \theta_1} \prod^{n-1}_{j=2}
{}_{j}\bar{P}_{l_j}^{l_{j-1}} (\theta_j) \, {}_{j}\bar{P}_{l'_j}^{l'_{j-1}} (\theta_j) \, d\Omega_{n-1} \,.
\end{align}
The $\theta_1$ integral is
\begin{align}\label{Orthogonality_Appendix_Theta1Eqn}
\int^{2\pi}_0 e^{-i(l_1 - l'_1)\theta_1} d\theta_1 = 2\pi \delta_{l_1 l'_1} \,.
\end{align}
The $\theta_j$ integral, for $2 \leq j \leq n-1$, is
\begin{align}\label{Orthogonality_Appendix_OtherThetaEqn}
\int^{\pi}_0 {}_{j}\bar{P}_{l_j}^{l_{j-1}} (\theta_j) \, {}_{j}\bar{P}_{l'_j}^{l'_{j-1}} (\theta_j) \, \sin^{j-1}\theta_j \, d\theta_j \,.
\end{align}
We first tackle the case in which we have at least one $l_j \neq l_j'$. Let $j$ be the smallest integer for which $l_j \neq l'_j$ occurs. If $j=1$, then by (\ref{Orthogonality_Appendix_Theta1Eqn}), (\ref{Orthogonality_Appendix_MasterEqn}) integrates to 0. If $2 \leq j \leq n-1$, because $j$ is the smallest integer such that $l_j \neq l'_j$ occurs, we have $l_{j-1} = l'_{j-1}$. We consider (\ref{Orthogonality_Appendix_OtherThetaEqn}) in the special case $l_{j-1} = l'_{j-1}$, which is shown in \citep{nDimSphericalHarmonics} to equal to
\begin{align}\label{Orthogonality_Appendix_OtherThetaEqn_SpCase}
& \int^{\pi}_0 {}_{j}\bar{P}_{l_j}^{l_{j-1}} (\theta_j) \, {}_{j}\bar{P}_{l'_j}^{l_{j-1}} (\theta_j) \, \sin^{j-1}\theta_j \, d\theta_j
= \delta_{l_j l'_j} \,.
\end{align}
%
Thus (\ref{Orthogonality_Appendix_OtherThetaEqn}) again integrates to 0, and orthogonality in (\ref{Orthogonality_Appendix}) is proven. Next, by imposing normality in (\ref{Orthogonality_Appendix}), we determine the normalization constant $A$. Consider the special case where $l_j = l'_j$ for all $1 \leq j \leq n-1$. By (\ref{Orthogonality_Appendix_Theta1Eqn}) and (\ref{Orthogonality_Appendix_OtherThetaEqn_SpCase}), (\ref{Orthogonality_Appendix_MasterEqn}) integrates to $|A|^2 2\pi$. We want this to normalize to 1, so
\begin{equation}
A = \frac{1}{\sqrt{2\pi}} \,.
\end{equation}
Thus, the $n$-dimensional spherical harmonics take the form
\begin{align}\label{nDimSphericalHarmonicsExplicitForm}
Y_{l_1,...,l_{n-1}} (\theta_1,...,\theta_{n-1}) = \frac{1}{\sqrt{2\pi}} e^{il_1 \theta_1}  \prod^{n-1}_{j=2} {}_{j}\bar{P}_{l_j}^{l_{j-1}} (\theta_j) \,,
\end{align}
and satisfy the orthonormality condition (\ref{Orthogonality_Appendix}).

Next we prove another useful identity
\begin{align}\label{completeness_Appendix}
& \sum_{\substack{l_1,...,\\ l_{n-1}}} Y_{l_1,...,l_{n-1}} (\theta_1,...,\theta_{n-1}) Y^*_{l_1,...,l_{n-1}} (\theta'_1,...,\theta'_{n-1}) \nonumber\\
& = \frac{1}{\sin^{n-2}\theta_{n-1}\sin^{n-3}\theta_{n-2} ... \sin\theta_{2}} \delta(\theta_{1}-\theta'_1)... \nonumber\\
& \cdot \delta(\theta_{n-1}-\theta'_{n-1})  \,.
\end{align}
This is a result of the completeness property of the spherical harmonics. By completeness, we can express any function $V(\theta_1,...,\theta_{n-1})$ on $S^{n-1}$ as a linear combination of $Y_{l_1,...,l_{n-1}} (\theta_1,...,\theta_{n-1})$:
\begin{align}\label{Lincomb}
V(\theta_1,...,\theta_{n-1}) = \sum_{\substack{l_1,...,\\ l_{n-1}}} V_{l_1,...,l_{n-1}} Y_{l_1,...,l_{n-1}} (\theta_1,...,\theta_{n-1}) \,.
\end{align}
Using the orthonormality condition (\ref{Orthogonality_Appendix}), we can determine the coefficients $V_{l_1,...,l_{n-1}}$ by
\begin{align}\label{Coef}
V_{l_1,...,l_{n-1}} = \int & Y^*_{l_1,...,l_{n-1}} (\theta_1,...,\theta_{n-1}) V(\theta_1,...,\theta_{n-1}) \nonumber\\
& d\Omega_{n-1} \,.
\end{align}
Substitute (\ref{Coef}) into (\ref{Lincomb}), and also by noting that
\begin{align}
d \Omega_{n-1} = \sin^{n-2}\theta_{n-1}\sin^{n-3}\theta_{n-2} ... \sin\theta_{2} d\theta_1 d\theta_2  ... d\theta_{n-1} \,,
\end{align}
we get
\begin{align}
& V(\theta_1,...,\theta_{n-1}) \nonumber\\
= & \sum_{\substack{l_1,...,\\ l_{n-1}}}
\left( \int Y^*_{l_1,...,l_{n-1}} (\theta'_1,...,\theta'_{n-1}) V(\theta'_1,...,\theta'_{n-1}) d\Omega'_{n-1} \right) \nonumber\\
& Y_{l_1,...,l_{n-1}} (\theta_1,...,\theta_{n-1}) \nonumber\\
= & \int \left\{ \sum_{\substack{l_1,...,\\ l_{n-1}}} Y^*_{l_1,...,l_{n-1}} (\theta'_1,...,\theta'_{n-1}) Y_{l_1,...,l_{n-1}} (\theta_1,...,\theta_{n-1}) \right\} \nonumber\\
& V(\theta'_1,...,\theta'_{n-1}) \sin^{n-2} \theta'_{n-1} \sin^{n-3} \theta'_{n-2}... \sin \theta'_{2} \nonumber\\
& d\theta'_{1}...d\theta'_{n-1} \,.
\end{align}
In order for this equation to hold, we must require the expression in the braces to equal to the expression on the right hand side of (\ref{completeness_Appendix}), thus proving (\ref{completeness_Appendix}).

Lastly, we shall prove
\begin{align}\label{UsefulIdentity_Appendix}
& \int Y_{l_1,...,l_{n-1}} (\theta_1,...,\theta_{n-1}) d\Omega_{n-1} \nonumber\\
= & \left(\frac{2\pi^{\frac{n}{2}}}{\Gamma(\frac{n}{2})}\right)^{\frac{1}{2}} \delta_{l_10}\delta_{l_20}...\delta_{l_{n-1}0} \,.
\end{align}
From (\ref{nDimSphericalHarmonicsExplicitForm}), we know that
\begin{equation}\label{Y0Form_Appendix}
Y_{0,...,0} (\theta_1,...,\theta_{n-1}) = \left( \frac{\Gamma(\frac{n}{2}) }{2\pi^{\frac{n}{2}}} \right)^{\frac{1}{2}}
\end{equation}
is a constant, where $\Gamma$ is the Gamma function. By the orthonormality condition (\ref{Orthogonality_Appendix}), it is the only $n$-dimensional spherical harmonic $Y_{l_1,...,l_{n-1}} (\theta_1,...,\theta_{n-1})$ that is a constant. Thus
\begin{align}
& \int Y_{0,...,0} (\theta_1,...,\theta_{n-1})\, d\Omega_{n-1} \nonumber\\
= & Y_{0,...,0}(\theta_1,...,\theta_{n-1})  \int d\Omega_{n-1}
= \left( \frac{\Gamma(\frac{n}{2}) }{2\pi^{\frac{n}{2}}} \right)^{\frac{1}{2}} \frac{2\pi^{\frac{n}{2}}}{\Gamma(\frac{n}{2})} \nonumber\\
= & \left( \frac{ 2\pi^{\frac{n}{2}} }{ \Gamma(\frac{n}{2})} \right)^{\frac{1}{2}} \,.
\end{align}
Next, consider the integral
\begin{align}
\int Y_{0,...,0}(\theta_1,...,\theta_{n-1}) Y_{l_1,...,l_{n-1}} (\theta_1,...,\theta_{n-1}) d\Omega_{n-1}
\end{align}
where $Y_{l_1,...,l_{n-1}} (\theta_1,...,\theta_{n-1})$ has at least one $l_j \neq 0$. By the orthogonality condition (\ref{Orthogonality_Appendix}), this integral equals 0. Thus we have
\begin{align}
0 & = \int Y_{0,...,0}(\theta_1,...,\theta_{n-1}) Y_{l_1,...,l_{n-1}} (\theta_1,...,\theta_{n-1}) d\Omega_{n-1} \nonumber\\
& = Y_{0,...,0}(\theta_1,...,\theta_{n-1}) \left\{ \int Y_{l_1,...,l_{n-1}} (\theta_1,...,\theta_{n-1}) d\Omega_{n-1} \right\}
\end{align}
Since $Y_{0,...,0}(\theta_1,...,\theta_{n-1})$ is a non-zero constant, the integral in the braces must equal to 0, completing the proof of (\ref{UsefulIdentity_Appendix}).

\section{Additional details on the derivations of the full-energy mapping CFT approach in the isotropic limit}\label{Appdendix:IsotropicDetailedDerivations}

In this section we provide some additional details on the derivations of the full-energy mapping CFT (FEMCFT) approach in the isotropic limit.

\subsection{Derivations of equations (7) and (8) of the main text}
First we note that $c_{\alpha, i, \mu}(\vec{p})$ in the Hamiltonians (2) and (4) of the main text are fermionic fields, so they satisfy the usual anticommutation relation
\begin{equation}
\{ c^{\dagger}_{\alpha, i, \mu}(\vec{p}), \, c_{\beta, j, \nu}(\vec{p}') \} = \delta_{\alpha\beta}\delta_{ij}\delta_{\mu \nu}\delta^{(n)}(\vec{p} - \vec{p}')
\end{equation}
in $n$ dimensions. We then expand $c_{\alpha, i, \mu}(\vec{p})$ as a linear combination of the $n-$dimensional spherical harmonics, which form a complete set of orthonormal functions on the $(n-1)$-sphere $S^{n-1}$. The expansion is given by equation (6) of the main text:
\begin{align}\label{VecInTermsOfMagnitude}
& c_{\alpha, i, \mu}(\vec{p}) = \frac{1}{p^{\frac{n-1}{2}}} \cdot \nonumber\\
& \sum_{l_1,...,l_{n-1}} Y_{l_1,...,l_{n-1}}(\theta_1,...,\theta_{n-1}) c_{l_1, ..., l_{n-1}, \alpha, i, \mu}(p) \,.
\end{align}
where $Y_{l_1,...,l_{n-1}} (\theta_1,...,\theta_{n-1})$ are the spherical harmonics in $n$ dimensions discussed in the previous section. Note that the coefficients in the expansion (\ref{VecInTermsOfMagnitude}) are only functions of $p \equiv |\vec{p}|$, and do not depend on any of the angular coordinates $\theta_1,...,\theta_{n-1}$. Also, a factor of $1/p^{\frac{n-1}{2}}$ has been factored out from each coefficient in the expansion, so that the remaining part of the coefficient, the fields $c_{l_1, ..., l_{n-1},\alpha, i, \mu}(p)$, satisfy the proper fermionic anticommutation relation
\begin{align}
& \{ c^{\dagger}_{l_1, ..., l_{n-1},\alpha, i, \mu}(p), \, c_{l'_1, ..., l'_{n-1},\beta, j, \nu}(p') \} \nonumber\\
= & \delta_{l_1l'_1}...\delta_{l_{n-1}l'_{n-1}}\delta_{\alpha\beta}\delta_{ij}\delta_{\mu \nu}\delta(p - p') \,.
\end{align}
Also, one can show that $c_{l_1, ..., l_{n-1},\alpha, i, \mu}(p)$  are related to $c_{\alpha, i, \mu}(\vec{p})$ by
\begin{align}\label{MagnitudeInTermsOfVec}
& c_{l_1, ..., l_{n-1},\alpha, i, \mu}(p) = \nonumber\\
& p^{\frac{n-1}{2}} \int d \Omega_{n-1} Y^*_{l_1,...,l_{n-1}} (\theta_1,...,\theta_{n-1}) c_{\alpha, i, \mu}(\vec{p}) \,,
\end{align}

We shall write the Hamiltonian in terms of these new fermionic fields $c_{l_1, ..., l_{n-1},\alpha, i, \mu}(p)$. By substituting (\ref{VecInTermsOfMagnitude}) into $H_I$ given by equation (4) of the main text, and by using $d^n \vec{p} = p^{n-1}dp d\Omega_{n-1}$ and (\ref{UsefulIdentity_Appendix}), $H_I$ can be written into the form of equation (7) of the main text:
\begin{align}\label{H_IBranchPt_Appendix}
H_{I} = \frac{\lambda \vec{S}}{\Gamma(\frac{n}{2})2^n\pi^{\frac{n}{2}}} \cdot \sum_{\mu, \nu} \int & dpdp' p^{\frac{n-1}{2}} p'^{\frac{n-1}{2}} \nonumber\\
& c^{\dagger}_{0,...,0,\alpha,i,\mu}(p) \vec{\sigma}_{\alpha\beta} c_{0,...,0,\beta,i,\nu}(p')\,.
\end{align}
This shows that the only partial waves that couple to the impurity $\vec{S}$ are those with $(l_1,...,l_{n-1}) = (0,...,0)$.

Similarly, substitute (\ref{VecInTermsOfMagnitude}) into $H_0$ given in equation (2) of the main text, and use the orthonormality condition (\ref{Orthogonality_Appendix}), we get
\begin{align}\label{IsotropicCase_PartialWaveOnlyCoupleAmongThemselves}
H_0 = \sum_{\substack{l_1,...,\\ l_{n-1}}} \int dp \, c^{\dagger}_{l_1, ..., l_{n-1}, \alpha, i, \mu}(p) c_{l_1, ..., l_{n-1}, \alpha, i, \mu}(p) \epsilon_{\mu} (p) \,.
\end{align}
Equation (\ref{H_IBranchPt_Appendix}) shows that the only partial waves that couple to the impurity $\vec{S}$ in $H_I$ are those with $(l_1,...,l_{n-1}) = (0,...,0)$, and equation (\ref{IsotropicCase_PartialWaveOnlyCoupleAmongThemselves}) shows that in the bath, there is only coupling among  partial waves of the same quantum numbers. Thus no partial waves with $\{l_1,...,l_{n-1}\} \neq \{0,...,0\}$ will couple to the impurity, either directly or indirectly. As a result, we only need to consider partial waves with $\{l_1,...,l_{n-1}\} = \{0,...,0\}$ in $H_0$. Thus $H_0$ reduces to equation (8) of the main text:
\begin{align}\label{H_0BranchPt_Appendix}
H_0 = \int dp \, c^{\dagger}_{0, ..., 0, \alpha, i, \mu}(p) c_{0, ..., 0, \alpha, i, \mu}(p) \epsilon_{\mu} (p) \,.
\end{align}

\subsection{Derivations of the anticommutation relations for the $c(\epsilon)$ and the $\Phi$ fields}
In this section, we show that the field $c_{l_1, ..., l_{n-1}, \alpha, i, \mu}(\epsilon_{\mu})$ defined in equation (9) of the main text and the field $\Phi_{0,...,0,\alpha,i}(\epsilon)$ defined in equation (10) of the main text are fermionic fields satisfying the proper anticommutation relations. Due to the delta function identity
\begin{equation}
\delta(\epsilon_{\mu} - \epsilon'_{\mu}) = \frac{\delta(p-p')}{\left| \frac{d\epsilon_{\mu}(p)}{dp}|_{p=p'} \right|} \,,
\end{equation}
the field $c_{l_1, ..., l_{n-1}, \alpha, i, \mu}(\epsilon_{\mu})$ obeys the anticommutation relation
\begin{align}
& \{ c^{\dagger}_{l_1, ..., l_{n-1}, \alpha, i, \mu}(\epsilon_{\mu}), \, c_{l'_1, ..., l'_{n-1}, \beta, j, \nu}(\epsilon'_{\nu}) \} \nonumber\\
= & \delta_{l_1l'_1}...\delta_{l_{n-1}l'_{n-1}}\delta_{\alpha\beta}\delta_{ij}\delta_{\mu \nu}\delta(\epsilon_{\mu} - \epsilon'_{\nu}) \,.
\end{align}
As a result, the composite field $\Phi_{0,...,0,\alpha,i}(\epsilon)$ satisfies the anticommutation relation
\begin{align}
\{ \Phi^{\dagger}_{0,...,0,\alpha,i}(\epsilon), \Phi_{0,...,0,\beta,j}(\epsilon') \} = \delta_{\alpha\beta}\delta_{ij}\delta(\epsilon - \epsilon')
\end{align}
as well.

\subsection{Writing the Hamiltonian in terms of the composite field $\Phi$}
In this section, we show the detailed derivations of equations (11) and (12) of the main text, i.e. how to write the Hamiltonian in terms of the composite fermion field $\Phi_{0,...,0,\alpha,i}(\epsilon)$. We tackle equation (11) of the main text first. By substituting equation (9) of the main text into equation (8) of the main text (i.e. equation (\ref{H_0BranchPt_Appendix}) in this appendix), and using equation (5) of the main text (WLOG assume $a_{1+}$ is positive), we get
\begin{align}
H_0 = & \int_0^{\infty}d\epsilon \, c^{\dagger}_{0, ..., 0, \alpha, i, +}(\epsilon) c_{0, ..., 0, \alpha, i, +}(\epsilon)\epsilon \nonumber\\
+ & \int_{-\infty}^0d\epsilon \, c^{\dagger}_{0, ..., 0, \alpha, i, -}(\epsilon) c_{0, ..., 0, \alpha, i, -}(\epsilon)\epsilon \,,
\end{align}
where we have relabelled the dummy variables $\epsilon_+ \rightarrow \epsilon$ and $\epsilon_- \rightarrow \epsilon$ in the first term and in the second term respectively. Then by using the definition of $\Phi_{0,...,0,\alpha,i}(\epsilon)$ given by equation (10) of the main text, $H_0$ becomes
\begin{align}
H_0 = \int_{-\infty}^{\infty} d\epsilon \, \Phi^{\dagger}_{0, ..., 0, \alpha, i}(\epsilon) \Phi_{0, ..., 0, \alpha, i}(\epsilon) \epsilon \,,
\end{align}
which is equation (11) of the main text.

As for $H_I$, by substituting equation (9) of the main text into equation (7) of the main text (i.e. equation (\ref{H_IBranchPt_Appendix}) in this appendix), we get
\begin{align}\label{H_IBeforeSpecialCondition}
& H_{I} = \frac{\lambda \vec{S}}{\Gamma(\frac{n}{2})2^n\pi^{\frac{n}{2}}} \cdot \sum_{\mu, \nu} \int d\epsilon_{\mu} \int d\epsilon'_{\nu} p^{\frac{n-1}{2}} p'^{\frac{n-1}{2}} \nonumber\\
& \left| \frac{d\epsilon_{\mu}(p)}{dp} \right|^{-\frac{1}{2}} \left| \frac{d\epsilon'_{\nu}(p')}{dp'} \right|^{-\frac{1}{2}}
c^{\dagger}_{0,...,0,\alpha,i,\mu}(\epsilon_{\mu}) \vec{\sigma}_{\alpha\beta} c_{0,...,0,\beta,i,\nu}(\epsilon'_{\nu})\,,
\end{align}
where
\begin{align}
\int d\epsilon_{\mu} & \equiv
\begin{cases}
\int_0^{\infty} d\epsilon_{+} & \text{if } \mu = + \\
\int_{-\infty}^0 d\epsilon_{-} & \text{if } \mu = -
\end{cases} \,, \nonumber\\
\int d\epsilon'_{\nu} & \equiv
\begin{cases}
\int_0^{\infty} d\epsilon'_{+} & \text{if } \nu = + \\
\int_{-\infty}^0 d\epsilon'_{-} & \text{if } \nu = -
\end{cases} \,.
\end{align}
Using equation (5) of the main text, $\left| \frac{d \epsilon_{\mu} (p)}{dp} \right|^{-\frac{1}{2}}$ and $p^{\frac{n-1}{2}}$ cancel off each other, and similarly $\left| \frac{d \epsilon'_{\nu} (p')}{dp'} \right|^{-\frac{1}{2}}$ and $p'^{\frac{n-1}{2}}$ cancel off each other. We get
\begin{align}
H_{I} = & \frac{\lambda \vec{S}}{an\Gamma(\frac{n}{2})2^n\pi^{\frac{n}{2}}} \nonumber\\
& \cdot \sum_{\mu, \nu} \int d\epsilon_{\mu} \int d\epsilon'_{\nu} c^{\dagger}_{0,...,0,\alpha,i,\mu}(\epsilon_{\mu}) \vec{\sigma}_{\alpha\beta} c_{0,...,0,\beta,i,\nu}(\epsilon'_{\nu})\,,
\end{align}
where $a \equiv |a_{1\mu}| = |a_{1\nu}|$. Writing out the sum over $\mu, \nu$ explicitly and relabelling the dummy variables $\epsilon_+ \rightarrow \epsilon$ and $\epsilon_- \rightarrow \epsilon$, we get
\begin{align}
& H_{I} =  \frac{\lambda \vec{S}}{an\Gamma(\frac{n}{2})2^n\pi^{\frac{n}{2}}} \nonumber\\
& \cdot \left\{ \int_0^{\infty} d\epsilon \int_0^{\infty} d\epsilon' c^{\dagger}_{0,...,0,\alpha,i,+}(\epsilon) \vec{\sigma}_{\alpha\beta} c_{0,...,0,\beta,i,+}(\epsilon') \right. \nonumber\\
& + \int_0^{\infty} d\epsilon \int_{-\infty}^0 d\epsilon' c^{\dagger}_{0,...,0,\alpha,i,+}(\epsilon) \vec{\sigma}_{\alpha\beta} c_{0,...,0,\beta,i,-}(\epsilon') \nonumber\\
& + \int_{-\infty}^0 d\epsilon \int_0^{\infty} d\epsilon' c^{\dagger}_{0,...,0,\alpha,i,-}(\epsilon) \vec{\sigma}_{\alpha\beta} c_{0,...,0,\beta,i,+}(\epsilon') \nonumber\\
& + \left. \int_{-\infty}^0 d\epsilon \int_{-\infty}^0 d\epsilon' c^{\dagger}_{0,...,0,\alpha,i,-}(\epsilon) \vec{\sigma}_{\alpha\beta} c_{0,...,0,\beta,i,-}(\epsilon') \right\} \nonumber\\
& = \frac{\lambda \vec{S}}{an\Gamma(\frac{n}{2})2^n\pi^{\frac{n}{2}}} \nonumber\\
& \cdot \int_{-\infty}^{\infty} d\epsilon \int_{-\infty}^{\infty} d\epsilon' \Phi^{\dagger}_{0,...,0,\alpha,i}(\epsilon) \vec{\sigma}_{\alpha\beta} \Phi_{0,...,0,\beta,i}(\epsilon') \,.
\end{align}
which is equation (12) of the main text.

\subsection{Derivation of equation (16) and (17) of the main text}
We first derive equation (17) of the main text. Let $\tau$ be the imaginary time $\tau \equiv it$, and define the complex plane $\mathbb{C}$ to which we map our KHOF model as equation (14) of the main text,
\begin{equation}
\mathbb{C} = \{ z \equiv \tau + ir \} \,.
\end{equation}
View the $\Psi_{\leftarrow / \rightarrow \alpha i}(r)$ fields defined in equation (13) of the main text in the Heisenberg picture, so that they now have time-dependence, and thus live on $\mathbb{C}$. Using equation (13) of the main text, we can see that $H_I$ given by equation (12) of the main text can be written in the form of equation (17) of the main text.

We now derive equation (16) of the main text. Using equation (13) of the main text, $H_0$ given by equation (11) of the main text can be written as
\begin{align}\label{r_nonnegative}
& H_0 = \frac{1}{2\pi} \int^{\infty}_0 dr \, \cdot \nonumber\\
& \left( \Psi^{\dagger}_{\leftarrow \alpha i}(r) i\frac{\partial}{\partial r} \Psi_{\leftarrow \alpha i}(r) - \Psi^{\dagger}_{\rightarrow \alpha i}(r) i\frac{\partial}{\partial r} \Psi_{\rightarrow \alpha i}(r) \right) \,.
\end{align}
We note that $0 \leq r \leq \infty$ in (\ref{r_nonnegative}). One can also show that $\Psi_{\leftarrow / \rightarrow \alpha i}(r)$ satisfy the anticommutation relations
\begin{align}\label{r_Field_anti-commutation}
\{\Psi^{\dagger}_{X \alpha i}(r) , \Psi_{X' \beta j}(r')\} = 2\pi \delta_{XX'} \delta_{\alpha \beta} \delta_{ij} \delta(r-r') \,.
\end{align}
Now equation (17) of the main text tells us that $H_I$ is only confined to the horizontal axis $r=0$. At $r \neq 0$, $H=H_0$ and we have the Heisenberg equations of motion
\begin{align}\label{HeisenbergEOM}
i\frac{\partial}{\partial t} \Psi_{X \alpha i}(\tau, r) = [\Psi_{X \alpha i}(\tau, r), H_0]
\end{align}
for each $X \in \{\leftarrow, \rightarrow\}$.
By substituting $H_0$ in the form of (\ref{r_nonnegative}) into the Heisenberg equation of motion (\ref{HeisenbergEOM}), and applying the anticommutation relation (\ref{r_Field_anti-commutation}) and the identity $[A, BC] = \{A, B\}C - B\{A, C\}$ for three arbitrary operators $A, B$ and $C$, the Heisenberg equation of motion reduces to
\begin{align}
\partial_{\bar{z}} \Psi_{\leftarrow \alpha i}(\tau, r) = 0 \,, \qquad \, \partial_z \Psi_{\rightarrow \alpha i}(\tau, r) = 0\,.
\end{align}
These are exactly the Cauchy-Riemann equations for holomorphic functions and antiholomorphic functions respectively, which imply that $\Psi_{\leftarrow \alpha i}(t, r)$ is a holomorphic function, and $\Psi_{\rightarrow \alpha i}(t, r)$ is an antiholomorphic function, on the upper half plane $\mathbb{C}_+ = \{z \equiv \tau + ir | r \geq 0\}$ (since in (\ref{r_nonnegative}), $r$ is non-negative, $\Psi_{\leftarrow/\rightarrow \alpha i} (\tau, r)$ live on the upper half plane). Equivalently, $\Psi_{\rightarrow \alpha i}(\tau, r)$ is a holomorphic function on the lower complex plane $\mathbb{C}_- = \{\bar{z} \equiv \tau - ir | r \geq 0\}$. Also, by the definition of $\Psi_{\leftarrow / \rightarrow \alpha i}(r)$ given by equation (13) of the main text, $\Psi_{\leftarrow \alpha i}(\tau, 0) = \Psi_{\rightarrow \alpha i}(\tau, 0)$, i.e. the two holomorphic functions $\Psi_{\leftarrow \alpha i}(\tau, r)$ on the upper complex plane and $\Psi_{\rightarrow \alpha i}(\tau, r)$ on the lower complex plane agree on the horizontal $\tau$-axis $r=0$. Thus they are analytic continuations of each other to the entire complex plane. This fact allow us to eliminate $\Psi_{\rightarrow \alpha i}(\tau, r)$ in terms of $\Psi_{\leftarrow \alpha i}(\tau, r)$ because the former is simply the analytic continuation of the latter into the lower complex plane:
\begin{equation}
\Psi_{\rightarrow \alpha i}(\tau, r) = \Psi_{\leftarrow \alpha i}(\tau, -r) \,.
\end{equation}
Thus (\ref{r_nonnegative}) becomes
\begin{align}\label{H0_Final_Appendix}
H_0 = \frac{1}{2\pi} \int^{\infty}_{-\infty} dr \left( \Psi^{\dagger}_{\leftarrow \alpha i}(\tau, r) i\frac{\partial}{\partial r} \Psi_{\leftarrow \alpha i}(\tau, r) \right) \,,
\end{align}
which is equation (16) of the main text.

\section{Additional details on the applications of the FEMCFT method in anisotropic materials}\label{Appendix:Anisotropic}
In this section, we provide additional details on the applications of our FEMCFT method to anisotropic KHOF models. When we describe our general method, we shall keep the dimension $n$ of our KHOF model to be arbitrary. From time to time, we illustrate our procedure with the cubic fermion system in three dimensions, whose dispersion is given by equation (21) of the main text. At such points we shall let $n=3$.

We have considered the FEMCFT approach in the isotropic limit, in which the dispersion relation of our KHOF system is isotropic. We now consider the changes that a KHOF system with an anisotropic dispersion relation brings about. In terms of the Hamiltonian $H=H_0+H_I$, $H_I$ remains unchanged, since the dispersion relation does not enter the expression of $H_I$. However, now in $H_0$, we not only have to expand $c_{\alpha, i, \mu}(\vec{p})$ as a linear combination of the spherical harmonics according to (\ref{VecInTermsOfMagnitude}), but also need to do so for $\epsilon_{\mu}(\vec{p})$ as well:
\begin{equation}\label{ExpandDispersionRelation}
\epsilon_{\mu}(\vec{p}) = \sum_{l_1,...,l_{n-1}} Y_{l_1,...,l_{n-1}}(\theta_1,...,\theta_{n-1}) \epsilon_{l_1,...,l_{n-1}, \mu}(p) \,.
\end{equation}
Substitute (\ref{VecInTermsOfMagnitude}), (\ref{ExpandDispersionRelation}) into equation (2) of the main text, we get
\begin{align}\label{AnisotropicH0}
H_0 = \sum_{L,L',L''}V_{L,L',L''} \int dp \, c^{\dagger}_{L, \alpha, i, \mu}(p) c_{L', \alpha, i, \mu}(p) \epsilon_{L'', \mu} (p) \,,
\end{align}
where for brevity we used the multi-index notation $L \equiv (l_1,...,l_{n-1}), L' \equiv (l'_1,...,l'_{n-1}), L'' \equiv (l''_1,...,l''_{n-1})$. The coupling strengths $V_{L,L',L''}$ are given by the hopping integral
\begin{align}\label{DefinitionOfCouplingConst}
& V_{L,L',L''} = \nonumber\\
& \int d\Omega_{n-1} Y^*_L(\theta_1,...,\theta_{n-1})Y_{L'}(\theta_1,...,\theta_{n-1})Y_{L''}(\theta_1,...,\theta_{n-1})\,.
\end{align}
We note that $V_{L,L',L''}$ depends on the dispersion relation $\epsilon_{\mu}(\vec{p})$ of the system, because different $\epsilon_{\mu}(\vec{p})$ observes different expansions in terms of spherical harmonics (\ref{ExpandDispersionRelation}), resulting in different sets of $Y_{L''}(\theta_1,...,\theta_{n-1})$ appearing in (\ref{DefinitionOfCouplingConst}). As a concrete example, we calculate $V_{L,L',L''}$ for the dispersion relation given by equation (21) of the main text in 3 dimensions,
\begin{equation}\label{SampleDispersionRelation_WithSign}
\epsilon_{\mu}(\vec{p}) = a_{1\mu}p^3 + a_{2\mu}p_1 \,,
\end{equation}
where $a_{1\mu}, a_{2\mu}$ are real constants satisfying
\begin{align}\label{DispersionConstantSignProperty}
a_{1+} = -a_{1-} \qquad \text{and} \qquad a_{2+} = -a_{2-} \,,
\end{align}
so that
\begin{equation}
\epsilon_{+}(\vec{p}) = -\epsilon_{-}(\vec{p})\,.
\end{equation}

WLOG align $p_1$ in the $z$-direction, so $p_1 = p_z = p\cos \theta$. Thus (\ref{SampleDispersionRelation_WithSign}) reads
\begin{equation}
\epsilon_{\mu}(\vec{p}) = a_{1\mu}p^3+a_{2\mu}p\cos \theta\,,
\end{equation}
whose expansion (\ref{ExpandDispersionRelation}) in terms of spherical harmonics in 3 dimensions is

\begin{equation}\label{ExpandDispersionRelation_OurExample}
\epsilon_{\mu}(\vec{p}) = a_{1\mu}p^3 \sqrt{4\pi}Y^0_0(\theta, \phi) + a_{2\mu}p \sqrt{ 4\pi/3 } Y^0_1(\theta, \phi) \,.
\end{equation}
This shows that for 3-dimensional systems with dispersion relation (\ref{SampleDispersionRelation_WithSign}), the only non-zero spherical harmonics in the expansion (\ref{ExpandDispersionRelation}) are $Y^0_0$ and $Y^0_1$, which enter (\ref{DefinitionOfCouplingConst}) as $Y_{L''}$. Thus the only non-zero $V_{L,L',L''}$ are $V_{L,L',(00)}$ and $V_{L,L',(01)}$. They are given by

\begin{equation}\label{HoppingStrength1stTerm}
V_{L,L',(00)} = \int d\Omega_2 Y^{m*}_l(\theta, \phi) Y^{m'}_{l'}(\theta, \phi) Y^{0}_{0}(\theta, \phi) = \frac{\delta_{ll'}\delta_{mm'}}{\sqrt{4\pi}} \,,
\end{equation}
where we have used $Y^{0}_{0}(\theta, \phi) = 1/\sqrt{4\pi}$ and the orthogonality condition (\ref{Orthogonality_Appendix}), and

\begin{align}\label{HoppingStrength2ndTermGeneral}
& V_{L,L',(01)} = \int d\Omega_2 Y^{m*}_l(\theta, \phi) Y^{m'}_{l'}(\theta, \phi) Y^{0}_{1}(\theta, \phi) \nonumber\\
= & \sqrt{ \frac{3(l-|m|)!(l'-|m|)!}{4\pi(2l+1)(2l'+1)(l+|m|)!(l'+|m|)!} } \frac{(l'+m)!}{(l'-m)!} \nonumber\\
& \cdot \{(l'-m)\delta_{l+1,l'}+(l+m)\delta_{l-1,l'}\} \delta_{mm'} \,,
\end{align}
where we have used
\begin{equation}
\int_0^{2\pi}d\phi \, e^{-i(m-m')\phi} = 2\pi \delta_{mm'}
\end{equation}
in the $\phi$-integral, the orthogonality relation
\begin{align}
\int_0^\pi P^m_l(\cos \theta)P^m_{l'}(\cos \theta)\sin\theta d\theta = \frac{2(l+m)!}{(2l+1)(l-m)!}\delta_{ll'}
\end{align}
and the recurrence formula
\begin{align}
& (2l+1)\cos\theta P^m_l(\cos \theta) \nonumber\\
 = & (l+m)P^m_{l-1}(\cos \theta) + (l-m+1)P^m_{l+1}(\cos \theta)
\end{align}
for the associated Legendre functions of the first kind, $P^m_l(\cos \theta)$, in the $\theta$-integral.

The fact that the only non-zero coupling strengths $V_{L,L',L''}$ are (\ref{HoppingStrength1stTerm}) and (\ref{HoppingStrength2ndTermGeneral}) gives rise to three interesting characteristics in our model. (i) The only partial waves that couple to the impurity are those with $m=0$. We have showed in  (\ref{H_IBranchPt}) that the impurity only couples to the $(m,l) = (0,0)$ partial wave. Anisotropy does not change this fact, since anisotropy does not alter $H_I$, as previously mentioned. Now (\ref{HoppingStrength1stTerm}) and (\ref{HoppingStrength2ndTermGeneral}) both have a factor of $\delta_{mm'}$, implying that there is no coupling between partial waves with $m \neq m'$. Thus the $(m,l) = (0,0)$ partial wave that couples to the impurity will only couple to other partial waves with $m=0$. We can hence ignore all partial waves with $m \neq 0$ from $H_0$. We note in particular that this a characteristic universal to all Hermitian systems in any dimension, not only unique to our example. This is because for Hermitian systems, $\epsilon(\vec{p})$ is real, so it can be expanded in terms of real spherical harmonics. Thus all $Y_{L''}$ in the hopping integral (\ref{DefinitionOfCouplingConst}) are real, and are independent of $\theta_1$: indeed, from (\ref{nDimSphericalHarmonicsExplicitForm}), we see that real spherical harmonics have no $\theta_1$ dependence. Thus the $\theta_1$-integral in (\ref{DefinitionOfCouplingConst}) only involve $Y^*_L$ and $Y_{L'}$, and equals to $\int_0^{2\pi}d\theta_1 e^{-i(l_1-l'_1)\theta_1} = 2\pi \delta_{l_1l'_1}$, which forbids coupling between partial waves with different $l_1$, which is the analogue of the quantum number $m$ in three dimensions. (ii) There is only nearest-neighbor hopping. This can be seen from the factor in the braces of (\ref{HoppingStrength2ndTermGeneral}): $\delta_{l+1,l'}$ and $\delta_{l-1,l'}$ implies hopping can only occur between partial waves whose $l$ and $l'$ differ by 1, resulting in only nearest-neighbor hopping. (iii) The nearest-neighbor hopping strength in (ii) decays with increasing $l$. Due to (i), we are only interested in the $m=0$ partial waves. Substitute $m=0$ into (\ref{HoppingStrength2ndTermGeneral}), we obtain

\begin{equation}
V_{(0l),(0l'),(01)} = \sqrt{\frac{3}{4\pi (2l+1)(2l'+1)}}(l'\delta_{l+1,l'}+l\delta_{l-1,l'}) \,.
\end{equation}
In other words, the coupling between the adjacent $(m=0, l)$-th and $(m=0, l+1)$-th partial waves is given by

\begin{equation}
V_{(0l),(0,l+1),(01)} = \sqrt{\frac{3}{4\pi (2l+1)(2l+3)}}(l+1) \,.
\end{equation}
By computing its derivative with respect to $l$, we see that the coupling strength $V_{(0l),(0,l+1),(01)}$ monotonically decreases as $l$ increases, for all $l \geq 0$.

The three properties (i), (ii), (iii) discussed above are depicted pictorially in the left picture of Figure 2 in the main text. In particular, property (iii) allows us to truncate the chain of partial waves in the figure, since the coupling strength decays as $l$ increases.

It is instructive to express the general $H_0$ in (\ref{AnisotropicH0}) into a matrix form in the partial wave basis, as shown in equation (18) of the main text:
\begin{equation}\label{H0MatrixForm_Appendix}
H_0 = \int dp \, C^{\dagger}_{\alpha , i, \mu}(p) A_{\mu}(p) C_{\alpha , i, \mu}(p) \,,
\end{equation}
where $A_{\mu}(p)$ is a matrix whose $(L, L')-$th element, $A_{L,L',\mu}(p)$, is given by
\begin{align}\label{MatrixElementDefinition}
& A_{L,L',\mu}(p) = \sum_{L''}V_{L,L',L''} \epsilon_{L'' \mu}(p) \nonumber\\
= & \int d\Omega_{n-1}Y^*_L(\theta_1,...,\theta_{n-1}) Y_{L'}(\theta_1,...,\theta_{n-1}) \epsilon_{\mu}(\vec{p}) \,,
\end{align}
$C_{\alpha , i, \mu}(p)$ is a column vector whose $L-$th element is given by $c_{L, \alpha , i, \mu}(p)$, and $C^{\dagger}_{\alpha , i, \mu}(p)$ is its Hermitian conjugate. Recall that property (i), which holds for all Hermitian systems, allows us to only include partial waves with $l_1 = 0$ in $H_0$. Thus here, the multi-indices $L$ and $L'$ only includes $(l_1= 0, l_2,...,l_{n-1})$. In particular, in 3 dimensions, we have the property that the multi-indices $L$ and $L'$ only includes $(m=0, l) \rightarrow l$, i.e. in 3 dimensions the multi-indices $L$ and $L'$ reduce to the single indices $l$ and $l'$: $L=l$, $L'=l'$.

In our specific example (\ref{SampleDispersionRelation_WithSign}), $A_{\mu}(p)$ is the tridiagonal matrix:
\begin{align}\label{PartitionedTridiagonalMatrix}
& A_{\mu}(p) =
  \left(\begin{array}{@{}c|ccccc@{}}
    a_{1\mu}p^3 & \frac{a_{2\mu}p}{\sqrt{3}} & & & & \\\hline
    \frac{a_{2\mu}p}{\sqrt{3}} & a_{1\mu}p^3 & \frac{2a_{2\mu}p}{\sqrt{15}} & & & \\
     & \frac{2a_{2\mu}p}{\sqrt{15}} & a_{1\mu}p^3 & \frac{3a_{2\mu}p}{\sqrt{35}} & & \\
     & & \frac{3a_{2\mu}p}{\sqrt{35}} & a_{1\mu}p^3 & \frac{4a_{2\mu}p}{\sqrt{63}} & \\
     & & & \ddots & \ddots & \ddots
  \end{array}\right) \,,
\end{align}
where the partitioning denoted by the horizontal and vertical lines will be explained shortly. We have $A_+(p) = -A_-(p)$ due to (\ref{DispersionConstantSignProperty}). In general, $A_{\mu}(p)$ is a $\infty \times \infty$ matrix, but by property (iii), since the hopping strength decays as the row/column number $L=l$ increases, in practice we can truncate it to an appropriate size for further calculations.

We next partition $A_{\mu}(p)$ as shown in (\ref{PartitionedTridiagonalMatrix}): the first block consists of only the (1,1)-th element, the second block consists of the remaining $(\infty-1) \times (\infty-1)$ square matrix, which we denote by $M_{\mu}(p)$, and the third and fourth parts consist of the $(\infty-1) \times 1$ column vector, denoted by $N_{\mu}(p)$, and its transpose, the $1 \times (\infty-1)$ row vector $N^{T}_{\mu}(p)$ respectively. This corresponds to separating the matrix form of $H_0$ in (\ref{H0MatrixForm}) into four terms:
\begin{align}\label{H0SeparateInto4Terms}
H_0 = \int dp & \left( \frac{\Gamma(n/2)}{2\pi^{\frac{n}{2}}} c^{\dagger}_{0, ..., 0, \alpha, i, \mu}(p) c_{0, ..., 0, \alpha, i, \mu}(p) \epsilon_{\mu}(p) \right. \nonumber\\
& + \mathcal{C}^{\dagger}_{\alpha,i,\mu}(p) M_{\mu}(p) \mathcal{C}_{\alpha,i,\mu}(p) \nonumber\\
& + \mathcal{C}^{\dagger}_{\alpha,i,\mu}(p)N_{\mu}(p)c_{0,...,0,\alpha,i,\mu}(p) \nonumber\\
& + \left. c^{\dagger}_{0,...,0,\alpha,i,\mu}(p) N^{T}_{\mu}(p) \mathcal{C}_{\alpha,i,\mu}(p) \right)\,.
\end{align}
We shall now formally define the objects that appear in (\ref{H0SeparateInto4Terms}). First of all, define the isotropic \textit{kernel} $\epsilon_{\mu}(p)$ of the dispersion relation $\epsilon_{\mu}(\vec{p})$ as
\begin{align}\label{KernelDefinition}
\epsilon_{\mu}(p) \equiv \int d \Omega_{n-1} \epsilon_{\mu}(\vec{p}) \,.
\end{align}
$\epsilon_{\mu}(p)$ can be seen as the ``isotropic part" of the dispersion relation. Next, let $\mathcal{L} \equiv (\l_1,...,l_{n-1}) \neq (0,...,0)$ be the multi-index obtained from $L$ by excluding $L=(0,...,0)$. $\mathcal{C}_{\alpha,i,\mu}(p)$ is the column vector whose $\mathcal{L}-$th element is given by $c_{{\mathcal{L}}, \alpha , i, \mu}(p)$, i.e. $\mathcal{C}_{\alpha,i,\mu}(p)$ is obtained from $C_{\alpha , i, \mu}(p)$ by excluding its first entry $c_{0, ..., 0, \alpha, i, \mu}(p)$. $\mathcal{C}^{\dagger}_{\alpha,i,\mu}(p)$ is the Hermitian conjugate of $\mathcal{C}_{\alpha,i,\mu}(p)$, i.e. $\mathcal{C}^{\dagger}_{\alpha,i,\mu}(p)$ is the row vector obtained from $C^{\dagger}_{\alpha , i, \mu}(p)$ by excluding its first entry $c^{\dagger}_{0, ..., 0, \alpha, i, \mu}(p)$. $M_{\mu}(p)$ is the matrix whose $(\mathcal{L}, \mathcal{L'})-$th element $M_{\mathcal{L}, \mathcal{L'}, \mu}(p)$ is defined by
\begin{equation}
M_{\mathcal{L}, \mathcal{L'}, \mu}(p) \equiv A_{\mathcal{L}, \mathcal{L'},\mu}(p) \,,
\end{equation}
i.e. $M_{\mu}(p)$ is the matrix obtained from $A_{\mu}(p)$ by removing its first row and first column, which corresponds to $L=(0,...,0)$ and $L'=(0,...,0)$ respectively. In our example, $M_{\mu}(p)$ is the bottom right block matrix in (\ref{PartitionedTridiagonalMatrix}). $N_{\mu}(p)$ is the column vector whose $\mathcal{L}-$th entry $N_{\mathcal{L},\mu}(p)$ is given by
\begin{align}
& N_{\mathcal{L}, \mu}(p) \equiv A_{\mathcal{L},(0,...,0),\mu}(p) \nonumber\\
= & \int d\Omega_{n-1}Y^*_{\mathcal{L}}(\theta_1,...,\theta_{n-1}) Y_{0,...,0}(\theta_1,...,\theta_{n-1}) \epsilon_{\mu}(\vec{p}) \nonumber\\
= & \left( \frac{\Gamma(n/2)}{2\pi^{\frac{n}{2}}} \right)^{\frac{1}{2}}\int d\Omega_{n-1} Y^*_{\mathcal{L}}(\theta_1,...,\theta_{n-1}) \epsilon_{\mu}(\vec{p}) \nonumber\\
= & \left( \frac{\Gamma(n/2)}{2\pi^{\frac{n}{2}}} \right)^{\frac{1}{2}}\int d\Omega_{n-1} Y^*_{\mathcal{L}}(\theta_1,...,\theta_{n-1})  \nonumber\\
& \sum_{L'} Y_{L'}(\theta_1,...,\theta_{n-1}) \epsilon_{L', \mu}(p) \nonumber\\
= & \left( \frac{\Gamma(n/2)}{2\pi^{\frac{n}{2}}} \right)^{\frac{1}{2}} \epsilon_{\mathcal{L}, \mu}(p) \,,
\end{align}
where in the third equality we have used the fact that $Y_{0,...0}(\theta_1,...,\theta_{n-1})$ is a real constant given by (\ref{Y0Form_Appendix}), in the fourth equality we have expanded the dispersion relation $\epsilon_{\mu}(\vec{p})$ in terms of the spherical harmonics (\ref{ExpandDispersionRelation}), and in the last equality we have used the orthogonality relation (\ref{Orthogonality_Appendix}). Recall that $\epsilon_{\mathcal{L}, \mu}(p)$ does not include $\epsilon_{0,...,0, \mu}(p)$. In our 3-dimensional example, the only non-zero $\epsilon_{\mathcal{L}, \mu}(p)$ is $\epsilon_{0,1, \mu}(p) = a_{2\mu}p\sqrt{4\pi/3}$, according to (\ref{ExpandDispersionRelation_OurExample}). Thus for our example
\begin{align}\label{N_OurExample}
N_{\mu}(p) = \left( \frac{\Gamma(\frac{3}{2})}{2\pi^{\frac{3}{2}}} \right)^{\frac{1}{2}}
\begin{pmatrix}
    \frac{a_{2\mu}p\sqrt{4\pi}}{\sqrt{3}}\\
    \vdots\\
    \text{\huge0}\\
    \vdots\\
  \end{pmatrix}
=
\begin{pmatrix}
    \frac{a_{2\mu}p}{\sqrt{3}}\\
    \vdots\\
    \text{\huge0}\\
    \vdots\\
  \end{pmatrix}
\,,
\end{align}
as can be seen from the first column of the matrix (\ref{PartitionedTridiagonalMatrix}), excluding its first entry. $N^T_{\mu}(p)$ is the transpose pf $N_{\mu}(p)$. Also in the derivation of the first term of (\ref{H0SeparateInto4Terms}) we have used
\begin{align}
& \int dp \, c^{\dagger}_{0, ..., 0, \alpha, i, \mu}(p) A_{(0,...,0),(0,...,0),\mu}(p) c_{0, ..., 0, \alpha, i, \mu}(p) \nonumber\\
= & \int dp \, c^{\dagger}_{0, ..., 0, \alpha, i, \mu}(p)  c_{0, ..., 0, \alpha, i, \mu}(p) \nonumber\\
& \cdot \int d\Omega_{n-1}Y^*_{0,...,0}(\theta_1,...,\theta_{n-1}) Y_{0,...0}(\theta_1,...,\theta_{n-1}) \epsilon_{\mu}(\vec{p}) \nonumber\\
= & \frac{\Gamma(n/2)}{2\pi^{\frac{n}{2}}} \int dp \, c^{\dagger}_{0, ..., 0, \alpha, i, \mu}(p)  c_{0, ..., 0, \alpha, i, \mu}(p) \int d\Omega_{n-1} \epsilon_{\mu}(\vec{p}) \nonumber\\
= & \frac{\Gamma(n/2)}{2\pi^{\frac{n}{2}}} \int dp \, c^{\dagger}_{0, ..., 0, \alpha, i, \mu}(p)  c_{0, ..., 0, \alpha, i, \mu}(p) \epsilon_{\mu}(p) \,, \nonumber\\
\end{align}
where in the last equality, we have used the definition of the kernel $\epsilon_{\mu}(p)$ in (\ref{KernelDefinition}).

Let us define the first term of $H_0$ in (\ref{H0SeparateInto4Terms}) to be $H_{00}$, and define the sum of the remaining three terms of $H_0$ in (\ref{H0SeparateInto4Terms}) to be $\Sigma_0$:
\begin{equation}
H_0 = H_{00} + \Sigma_0 \,,
\end{equation}
\begin{equation}\label{H_00DDefinition}
H_{00} \equiv \frac{\Gamma(n/2)}{2\pi^{\frac{n}{2}}} \int dp \, c^{\dagger}_{0, ..., 0, \alpha, i, \mu}(p) c_{0, ..., 0, \alpha, i, \mu}(p) \epsilon_{\mu} (p) \,,
\end{equation}
\begin{align}
\Sigma_0 \equiv \int dp & \left( \mathcal{C}^{\dagger}_{\alpha,i,\mu}(p) M_{\mu}(p) \mathcal{C}_{\alpha,i,\mu}(p) \right. \nonumber\\
& + \mathcal{C}^{\dagger}_{\alpha,i,\mu}(p)N_{\mu}(p)c_{0,...,0,\alpha,i,\mu}(p) \nonumber\\
& + \left. c^{\dagger}_{0,...,0,\alpha,i,\mu}(p) N^{T}_{\mu}(p) \mathcal{C}_{\alpha,i,\mu}(p) \right)\,.
\end{align}
We make these definitions because $H_{00}$ is the ``isotropic part" of $H_0$, whereas $\Sigma_0$ is the ``anisotropic part" of $H_0$. Indeed, $H_{00}$ only includes the $L=(0,...,0)$ partial wave, and $\Sigma_0$ contains the remaining $\mathcal{L}\neq (0,...,0)$ partial waves. For systems with isotropic dispersion relations,  only the $L=(0,...,0)$ partial wave couples to the impurity, and we can ignore all $\mathcal{L}\neq 0$ partial waves, so $\Sigma_0 = 0$ and $H_0 = H_{00}$. All contributions to the Hamiltonian due to anisotropy of the system are captured in $\Sigma_0$.

We shall now write the partition function $\mathcal{Z}$ for our system. Let $\psi_{L,\alpha,i,\mu}(p)$ be the Grassmann variable obtained from the operator $c_{L,\alpha,i,\mu}(p)$ acting on a coherent state. We have
\begin{align}\label{Action_Into2Factors}
& \mathcal{Z} = \int D(\bar{\psi}, \psi) e^{- S[\bar{\psi}, \psi]} \nonumber\\
= & \int D(\bar{\psi_0}, \psi_0) e^{- S[\bar{\psi_0}, \psi_0]} \int D(\bar{\psi_{\mathcal{L}}}, \psi_{\mathcal{L}}) e^{- S[\bar{\psi_{\mathcal{L}}}, \psi_{\mathcal{L}}]}\,,  \nonumber\\
\end{align}
where $S[\bar{\psi_0}, \psi_0]$ is the part of the action that only involves the $L=(0,...,0)$ partial wave, and $S[\bar{\psi_{\mathcal{L}}}, \psi_{\mathcal{L}}]$ is the part of the action that involves the remaining $\mathcal{L} \neq (0,...,0)$ partial waves. As previously discussed, in our Hamiltonian $H = H_{00} + \Sigma_0 + H_I$, $H_{00}$ and $H_I$ only includes the $L=(0,...,0)$ partial wave, so $H_{00}$ and $H_I$ appears in $S[\bar{\psi_0}, \psi_0]$, whereas $\Sigma_0$ includes the remaining $\mathcal{L} \neq (0,...,0)$ partial waves, so $\Sigma_0$ appears in $S[\bar{\psi_{\mathcal{L}}}, \psi_{\mathcal{L}}]$. We thus have
\begin{align}\label{Action_Isotropic}
S[\bar{\psi_0}, \psi_0] = & \sum_{\alpha,\beta,i,\mu}\int d\tau \nonumber\\
& \left\{ \int dp \, \bar{\psi}_{0,...,0,\alpha,i,\mu}(p)\partial_{\tau}\psi_{0,...,0,\alpha,i,\mu}(p) \right. \nonumber\\
& + \int dp H_{00}(\bar{\psi}_{0,...,0,\alpha,i,\mu}(p), \psi_{0,...,0,\alpha,i,\mu}(p))\nonumber\\
& + \left. \int dp \, dp' H_{I}(\bar{\psi}_{0,...,0,\alpha,i,\mu}(p), \psi_{0,...,0,\beta,i,\mu}(p')) \right\} \nonumber\\
\end{align}
and
\begin{align}\label{Action_Anisotropic}
S[\bar{\psi_{\mathcal{L}}}, \psi_{\mathcal{L}}] = \sum_{\alpha,i,\mu}\int d\tau dp \, & \{ \bar{\mathcal{F}}_{\alpha,i,\mu}(p) (\partial_{\tau} + M_{\mu}(p)) \mathcal{F}_{\alpha,i,\mu}(p) \nonumber\\
& + \bar{\mathcal{F}}_{\alpha,i,\mu}(p)\mathcal{N}_{\alpha,i,\mu}(p) \nonumber\\
& + \bar{\mathcal{N}}^T_{\alpha,i,\mu}(p) \mathcal{F}_{\alpha,i,\mu}(p) \}\,, \nonumber\\
\end{align}
where $\mathcal{F}_{\alpha,i,\mu}(p)$ is the column vector whose $\mathcal{L}-$th element is given by $\psi_{{\mathcal{L}}, \alpha , i, \mu}(p)$, $\bar{\mathcal{F}}_{\alpha,i,\mu}(p)$ is the row vector whose $\mathcal{L}-$th element is given by $\bar{\psi}_{{\mathcal{L}}, \alpha , i, \mu}(p)$,
\begin{equation}
\mathcal{N}_{\alpha,i,\mu}(p) \equiv N_{\mu}(p) \psi_{0,...,0, \alpha, i, \mu}(p) \,,
\end{equation}
and
\begin{equation}
\bar{\mathcal{N}}^T_{\alpha,i,\mu}(p) \equiv \bar{ \psi }_{0,...,0, \alpha, i, \mu}(p) N^T_{\mu}(p)\,.
\end{equation}
We then integrate out the Grassmann field $\bar{\mathcal{F}}_{\alpha,i,\mu}(p) $, $\mathcal{F}_{\alpha,i,\mu}(p) $ in  (\ref{Action_Anisotropic}), after which the second factor of (\ref{Action_Into2Factors}) becomes
\begin{align} \label{Action_Anisotropicadd}
& \int D(\bar{\psi_{\mathcal{L}}}, \psi_{\mathcal{L}}) e^{- S[\bar{\psi_{\mathcal{L}}}, \psi_{\mathcal{L}}]} \nonumber\\
= & e^{\sum_{\alpha,i,\mu}\int d\tau dp \{\bar{\mathcal{N}}^T_{\alpha,i,\mu}(p)M^{-1}_{\mu}(p) \mathcal{N}_{\alpha,i,\mu}(p)\}} \nonumber\\
\equiv & e^{\sum_{\alpha,i,\mu}\int d\tau dp \{\bar{ \psi }_{0,...,0, \alpha, i, \mu}(p) N^T_{\mu}(p)M^{-1}_{\mu}(p) N_{\mu}(p) \psi_{0,...,0, \alpha, i, \mu}(p)\}}\,.
\end{align}
where we have taken the static limit,  $\partial_{\tau} \rightarrow -iw\rightarrow 0$, in  (\ref{Action_Anisotropic}). This is a good approximation that well captures the renormalization of the isotropic part in low-energy window, as long as the system described by $M_{\mu}(p)$ is gapped. Notice that the factor in Eq.  (\ref{Action_Anisotropicadd})  now only includes the $L=(0,...,0)$ partial wave. Thus the partition function (\ref{Action_Into2Factors}) becomes
\begin{align}
\mathcal{Z} = \int D(\bar{\psi_0}, \psi_0) e^{- (S[\bar{\psi_0}, \psi_0] + S_{\Sigma_0})}\,,
\end{align}
where
\begin{align}
S_{\Sigma_0} \equiv - \sum_{\alpha,i,\mu}\int d\tau dp & \left\{\bar{ \psi }_{0,...,0, \alpha, i, \mu}(p) N^T_{\mu}(p)M^{-1}_{\mu}(p) N_{\mu}(p) \right. \nonumber\\
& \left. \psi_{0,...,0, \alpha, i, \mu}(p) \right\}\,.
\end{align}
This means that $S_{\Sigma_0}$ contributes an additional term of
\begin{align}
-\int dp \, \bar{ \psi }_{0,...,0, \alpha, i, \mu}(p) N^T_{\mu}(p)M^{-1}_{\mu}(p) N_{\mu}(p) \psi_{0,...,0, \alpha, i, \mu}(p)
\end{align}
to the braced integrand of $S[\bar{\psi_0}, \psi_0]$ given in (\ref{Action_Isotropic}). In terms of the Hamiltonian, this term is
\begin{align}
-\int dp \, c^{\dagger}_{0,...,0, \alpha, i, \mu}(p) N^T_{\mu}(p)M^{-1}_{\mu}(p) N_{\mu}(p) c_{0,...,0, \alpha, i, \mu}(p)\,.
\end{align}
Thus we can write $H_0$ as
\begin{align}
& H_0  = H_{00} + \Sigma_0 \nonumber\\
= & \frac{\Gamma(n/2)}{2\pi^{\frac{n}{2}}} \int dp \, c^{\dagger}_{0, ..., 0, \alpha, i, \mu}(p) c_{0, ..., 0, \alpha, i, \mu}(p) \epsilon_{\mu} (p) \nonumber\\
& -\int dp \, c^{\dagger}_{0,...,0, \alpha, i, \mu}(p) N^T_{\mu}(p)M^{-1}_{\mu}(p) N_{\mu}(p) c_{0,...,0, \alpha, i, \mu}(p) \nonumber\\
\equiv & \frac{\Gamma(n/2)}{2\pi^{\frac{n}{2}}} \int dp \, c^{\dagger}_{0, ..., 0, \alpha, i, \mu}(p) c_{0, ..., 0, \alpha, i, \mu}(p) \{ \epsilon_{\mu} (p) + \epsilon_{\Sigma \mu} (p) \} \nonumber\\
\equiv & \int dp \, c^{\dagger}_{0, ..., 0, \alpha, i, \mu}(p) c_{0, ..., 0, \alpha, i, \mu}(p) \tilde{\epsilon}_{\mu}(p)\,,
\end{align}
where in the second equality we have used the definition of $H_{00}$ given in (\ref{H_00DDefinition}), in the third equality we have made the definition
\begin{equation}\label{EpsilonSigmaDefinition}
\epsilon_{\Sigma \mu} (p) \equiv -\frac{2\pi^{\frac{n}{2}}}{\Gamma(n/2)} N^T_{\mu}(p)M^{-1}_{\mu}(p) N_{\mu}(p)\,,
\end{equation}
and in the last equality we have made the definition
\begin{equation}\label{EffectiveDispersionRelationDefinition}
\tilde{\epsilon}_{\mu}(p) \equiv \frac{\Gamma(n/2)}{2\pi^{\frac{n}{2}}} \left(\epsilon_{\mu} (p) + \epsilon_{\Sigma \mu} (p)\right)\,.
\end{equation}
We call $\tilde{\epsilon}_{\mu}(p)$ the \textit{effective dispersion relation} of the system.

We note that the kernel $\epsilon$ satisfies the relation
\begin{equation}\label{KernelSignProperty}
\epsilon_+(p) = -\epsilon_-(p)\,,
\end{equation}
because
\begin{align}
\epsilon_+(p) \equiv \int d \Omega_{n-1} \epsilon_+(\vec{p}) = - \int d \Omega_{n-1} \epsilon_-(\vec{p}) \equiv -\epsilon_-(p)\,,
\end{align}
where in the first and last equalities we have used the definition of the kernel given in (\ref{KernelDefinition}), and in the second equality we have used the relation $\epsilon_+(\vec{p}) = -\epsilon_-(\vec{p})$. Similarly, $\epsilon_{\Sigma \mu} (p)$ satisfies the relation
\begin{equation}\label{EpsilonSigmaSignProperty}
\epsilon_{\Sigma +} (p) = -\epsilon_{\Sigma -} (p)\,.
\end{equation}
To see this, notice that in the definition of $\epsilon_{\Sigma \mu} (p)$ given in (\ref{EpsilonSigmaDefinition}), each of the factors $N^T_{\mu}(p), M^{-1}_{\mu}(p),$ and $N_{\mu}(p)$ flips a sign when $\mu$ flips a sign: indeed, the entries of $N^T_{\mu}(p), M_{\mu}(p)$ and $N_{\mu}(p)$ are just the entries of the matrix $A$ defined in (\ref{MatrixElementDefinition}), whose integrand contains a factor of $\epsilon_{\mu}(\vec{p})$, which flips a sign as $\mu$ flips a sign. By matrix inverse properties we also have
\begin{align}
M^{-1}_+(p) = (-M_-(p))^{-1} = -M^{-1}_-(p)\,,
\end{align}
completing the proof. Thus the effective dispersion relation $\tilde{\epsilon}_{\mu}(p)$ also satisfies the relation
\begin{equation}
\tilde{\epsilon}_{+}(p) = -\tilde{\epsilon}_{-}(p)\,,
\end{equation}
due to (\ref{EffectiveDispersionRelationDefinition}), (\ref{KernelSignProperty}) and (\ref{EpsilonSigmaSignProperty}).

Let us compute the effective dispersion relation $\tilde{\epsilon}_{\mu}(p)$ for our example. The kernel is given by
\begin{align}
\epsilon_{\mu}(p) & = \int d\Omega_2 \, \epsilon_{\mu}(\vec{p}) \nonumber\\
& = \int d\Omega_2 \, \left\{ a_{1\mu}(p_x^2+p_y^2+p_z^2)^{\frac{3}{2}} + a_{2\mu}p_z \right\} \nonumber\\
& = \int_0^{2\pi} d \phi \int_0^{\pi} \sin(\theta) d \theta \, \left\{ a_{1\mu}p^3 + a_{2\mu}p\cos(\theta) \right\} \nonumber\\
& = 4\pi a_{1\mu} p^3\,.
\end{align}
As for $\epsilon_{\Sigma \mu} (p)$ defined in (\ref{EpsilonSigmaDefinition}), since $N_{\mu}(p)$ in our example only has its first entry being non-zero, as seen from (\ref{N_OurExample}), $\epsilon_{\Sigma \mu} (p)$ becomes
\begin{align}
\epsilon_{\Sigma \mu} (p) = -\frac{4\pi}{3} a_{2\mu}^2p^2 \left( M^{-1}_{\mu}(p) \right)_{11} \,,
\end{align}
where $\left( M^{-1}_{\mu}(p) \right)_{11}$ denotes the (1,1)-th element of $M^{-1}_{\mu}(p)$. Recall that $M_{\mu}(p)$ is the lower right block matrix of $A_{\mu}(p)$ in (\ref{PartitionedTridiagonalMatrix}):
\begin{align}\label{MMatrix}
& M_{\mu}(p) =
  \left(\begin{array}{@{}ccccc@{}}
     a_{1\mu}p^3 & \frac{2a_{2\mu}p}{\sqrt{15}} & & & \\
     \frac{2a_{2\mu}p}{\sqrt{15}} & a_{1\mu}p^3 & \frac{3a_{2\mu}p}{\sqrt{35}} & & \\
     & \frac{3a_{2\mu}p}{\sqrt{35}} & a_{1\mu}p^3 & \frac{4a_{2\mu}p}{\sqrt{63}} & \\
     & & \ddots & \ddots & \ddots
  \end{array}\right) \,.
\end{align}
By the discussion below (\ref{PartitionedTridiagonalMatrix}), we may truncate $A_{\mu}(p)$, and thus $M_{\mu}(p)$, to a desired size for actual calculations. Let $M^r_{\mu}(p)$ denote the $r \times r$ truncated matrix of $M_{\mu}(p)$, i.e. $M^r_{\mu}(p)$ is obtained from $M_{\mu}(p)$ by taking only its first $r$ rows and first $r$ columns. Because $M^r_{\mu}(p)$ is a tridiagonal matrix, one can show that
\begin{equation}
\left( M^{r}_{\mu}(p) \right)^{-1}_{11} = \frac{G^r_{\mu}(p)}{D^r_{\mu}(p)} \,,
\end{equation}
where $D^r_{\mu}(p)$ is the determinant of $ M^{r}_{\mu}(p) $, which can be computed from the recurrence relation
\begin{align}
D^r_{\mu}(p) = a_{1\mu}p^3D^{r-1}_{\mu}(p) - \frac{(a_{2\mu}pr)^2}{(2r-1)(2r+1)}D^{r-2}_{\mu}(p)\,,
\end{align}
with $D^0_{\mu}(p)=1, D^1_{\mu}(p)=a_{1\mu}p^3$, and $G^r_{\mu}(p)$ is a polynomial in $p$ that satisfies the same recurrence relation
\begin{align}
G^r_{\mu}(p) = a_{1\mu}p^3G^{r-1}_{\mu}(p) - \frac{(a_{2\mu}pr)^2}{(2r-1)(2r+1)}G^{r-2}_{\mu}(p)\,,
\end{align}
but with different initial terms $G^0_{\mu}(p)=0, G^1_{\mu}(p)=1$. Thus in our example,
\begin{equation}\label{UnsubstitutedEffectiveDispersionRelation}
\tilde{\epsilon}_{\mu}(p) = a_{1\mu}p^3 - \frac{(a_{2\mu}p)^2}{3}\frac{G^r_{\mu}(p)}{D^r_{\mu}(p)}\,,
\end{equation}
where $r \rightarrow \infty$.

We view the term $\epsilon_{\Sigma \mu}(p)$, which captures all anisotropic effects, as a correction to the isotropic term of the dispersion relation, the kernel $\epsilon_{\mu}(p)$. We approximate $\epsilon_{\Sigma \mu}(p)$ by evaluating its value at $p_F$, before adding it to the kernel $\epsilon_{\mu}(p)$. $p_F$ satisfies the property that $\tilde{\epsilon}_{\mu}(p_F)=0$, i.e. $p_F$ is a non-negative real root of $\tilde{\epsilon}_{\mu}(p)$ ($p_F$ needs to be non-negative because it is a special value of $p \equiv |\vec{p}|$). In general, $\tilde{\epsilon}_{\mu}(p)$ may have no non-negative real root, in which case the system is gapped and we do not consider such gapped systems. Also, $\tilde{\epsilon}_{\mu}(p)$ may have more than one non-negative real root. In such cases we pick the non-negative real root of $\tilde{\epsilon}_{\mu}(p)$ that gives rise to the greatest density of states (DOS) to be the $p_F$ at which we evaluate $\epsilon_{\Sigma \mu}(p_F)$. In other words, we pick the non-negative real root that results in the least $|d \tilde{\epsilon}_{\mu}(p_F)/dp|$. If there exists more than one non-negative real root of $\tilde{\epsilon}_{\mu}(p)$ that give rise to the same greatest DOS, we evaluate $\epsilon_{\Sigma \mu}(p)$ at each of them - this will result in channel-multiplying.

In our example, $\tilde{\epsilon}_{\mu}(p)$ given by (\ref{UnsubstitutedEffectiveDispersionRelation}) has more than one non-negative real root, with the smallest of them giving rise to the least $|d \tilde{\epsilon}_{\mu}(p)/dp|$, thus the greatest DOS. Hence there is only one $p_F$ at which we evaluate $\epsilon_{\Sigma \mu}(p_F)$, namely this smallest non-negative real root of $\tilde{\epsilon}_{\mu}(p)$. For all even values of $r$, this smallest non-negative real root of $\tilde{\epsilon}_{\mu}(p)$ is 0. For odd values of $r$, as $r \rightarrow \infty$, the smallest non-negative real root also approaches 0. Thus for our example, $p_F=0$, $\epsilon_{\Sigma \mu}(0) = 0$, and our effective dispersion relation becomes
\begin{equation}\label{FinalEffectiveDispersionRelation}
\tilde{\epsilon}_{\mu}(p) = a_{1\mu}p^3\,.
\end{equation}
Note that the effective dispersion relation (\ref{FinalEffectiveDispersionRelation}) is in the form of equation (5) of the main text in three dimensions. As a result, all subsequent derivations following equation (5) of the main text hold, and our FEMCFT approach applies.

\end{document}